\documentclass[11pt,urlcolor=black, linkcolor=black]{article} 
\pdfoutput=1
\usepackage{cite}
\usepackage{amsmath, amsthm, amssymb,slashed, url}
\usepackage{ifpdf}
\ifpdf
  \usepackage[pdftex]{graphicx}
  \usepackage{epstopdf}
\else  
  \usepackage[dvips]{graphicx}
\fi

\usepackage[usenames,dvipsnames]{color}
\usepackage[colorlinks,citecolor=black]{hyperref}

\parskip=\baselineskip
\def\Bbb{\mathbb}
\def\Tr{{\rm Tr}}
\def\16{{\bf 16}}
\def\Pin{{\mathrm{Pin}}}
\def\ii{{\mathcal I}}
\def\1{{\bf 1}}
\def\2{{\bf 2}}
\def\3{{\bf 3}}
\def\4{{\bf 4}}

\def\x{{\sf x}}
\def\y{{\sf y}}
\def\pin{{\mathrm{pin}}}

\def\mod{{\mathrm{mod}}}
\def\I{{\mathrm I}}
\def\bar{\overline}
\def\I{{\mathfrak I}}
\def\II{{\mathcal I}}

\def\CP{\Bbb{CP}}
\def\a{{\sf a}}

\def\tr{{\mathrm{tr}}}

\def\h{\widehat}
\def\bar{\overline}

\def\ra{\rangle}

\def\bp{\begin{pmatrix}}
\def\ep{\end{pmatrix}}
\def\la{\langle}
\def\R{{\Bbb{R}}}
\def\Z{{\Bbb{Z}}}

\def\RR{{\mathcal R}}
\def\S{{\mathcal S}}

\def\Pf{{\mathrm{Pf}}}
\def\DD{\slashed{D}}
\numberwithin{equation}{section}

\def\d{\mathrm d}

\def\L{{\mathcal L}}
\def\C{{\Bbb C}}
\def\Z{{\Bbb Z}}
\def\bar{\overline}

\def\bar{\overline}
\begin{document}

\def\be{\begin{equation}}
\def\ee{\end{equation}}
  
\def\aa{{\sf a}}
\def\bb{{\sf b}}
\def\a{{\bf a}}
\def\x{{\bf x}}
\def\y{{\bf y}}
\def\z{{\bf z}}
\def\d{{\mathrm d}}
 \def\sym{{\mathrm{sym}}}
 \def\zotimes{{\otimes N}}
 \def\cl{{\mathrm{cl}}}
\def\h{\widehat}
\def\t{\widetilde}
\def\CC{{\mathcal C}}
\def\tr{{\mathrm{tr}}}
\def\d{{\mathrm{d}}}
\def\H{{\mathcal H}}
\def\O{{\mathcal O}}
\def\Tr{{\mathrm{Tr}}}
\def\diag{{\mathrm{diag}}}
\def\vt{\vartheta}
\def\i{{\mathrm i}}
\def\la{\langle}
\def\ra{\rangle}
\def\Re{{\mathrm{Re}}}
\def\Im{{\mathrm{Im}}}
\def\S{{\mathcal S}}
\def\CC{{\Bbb C}}
\def\C{{\mathcal C}}
\def\VolH{V_H}
\def\SOS{{{\mathrm{SO}}(3)\times {\mathrm{SU}}(2)}}
\def\U{{\mathrm U}}
\def\Spin{\mathrm{Spin}}
\def\spinc{\mathrm{spin}_c}
\def\SU{\mathrm{SU(2)}}
\def\spinSU{\mathrm{spin \text{-} SU(2)}}
 \begin{titlepage}
\begin{flushright}
%hep-th/yymm.nnnn
\end{flushright}
\vskip 1.5in
\begin{center}
{\bf\Large{A New SU(2) Anomaly}}
\vskip
0.5cm {Juven Wang,$^1$ Xiao-Gang Wen,$^2$  and Edward Witten$^1$} \\ \vskip.8cm {\small{ \textit{$^1$ School of
Natural Sciences, Institute for Advanced Study,}\vskip -.4cm
{\textit{Einstein Drive, Princeton, NJ 08540 USA}}} \vskip-.4cm {\small{ \textit{$^2$ {Department of Physics, Massachusetts Institute of Technology, }\vskip -.4cm
{\textit{Cambridge, MA 02139 USA}}}

}}}
\end{center}
\vskip 0.5in
\baselineskip 16pt
\abstract{A familiar anomaly affects SU(2) gauge theory in four dimensions:  a theory with an odd number of fermion multiplets in the spin 1/2 representation of the gauge
group, and more generally in representations of spin $2r+1/2$, is inconsistent.  We describe here a more subtle anomaly that  can affect 
 SU(2) gauge theory in four dimensions under the condition that fermions transform with half-integer spin under SU(2)  and bosons with integer spin.  Such a theory,
 formulated in a way that requires no choice of spin structure, 
and with an odd number of fermion multiplets in representations of spin $4r+3/2$,  is inconsistent. The theory is consistent if one picks a spin or $\spinc$ structure.
Under Higgsing to U(1), the new SU(2) anomaly reduces to a known
 anomaly of ``all-fermion electrodynamics.''   Like that theory, an SU(2) theory with an odd number of fermion multiplets in representations of spin $4r+3/2$ can provide
 a boundary state for a five-dimensional gapped theory whose partition function on a closed five-manifold $Y$ is $(-1)^{\int_Y w_2w_3}$.  All statements have analogs with SU(2) replaced by
 Sp(2N). There is also an analog in five dimensions. }
\date{January, 2018}
\end{titlepage}
\tableofcontents
\section{Introduction}\label{introduction}

An SU(2) gauge theory with an odd number of fermion multiplets that transform in the spin 1/2 representation of the gauge group is inconsistent \cite{Witten}.   More generally, fermion
multiplets that transform under SU(2) in the spin $2r+1/2$ representation -- for brevity we will call these multiplets of isospin $2r+1/2$ -- contribute to this anomaly: an SU(2) gauge
theory with an odd number of fermion multiplets of isospin $2r+1/2$ is inconsistent.  

This anomaly is most systematically described, as in \cite{Witten},  using the fact that $\pi_4(\SU)=\Z_2$, and the relation of this to the
mod 2 index of the Dirac operator in five dimensions.  However, as we will
explain in section \ref{Two}, there is a more elementary way to see the anomaly using the fact that in SU(2) gauge theory, the number of fermion zero-modes in an instanton field can 
be odd.\footnote{This important property of SU(2) gauge theory was pointed out by S. Coleman at Aspen in the summer of 1976.}

This familiar SU(2) anomaly receives contributions from fermion multiplets of isospin $2r+1/2$ but not from multiplets in other representations.   However, in the present paper, we will describe
a similar but more subtle  anomaly in SU(2) gauge theory that receives contributions from (and only from) fermion multiplets of isospin $4r+3/2$.     The new anomaly depends on a refinement of
what one might mean by SU(2) gauge theory.   We consider an SU(2) gauge theory in which all fermion fields have half-integral isospin and all boson fields have integral isospin.  Such a theory
has no gauge-invariant local fields that are fermionic, so as in \cite{BiS}, one can view it as a possible critical point in a theory that microscopically is constructed from bosons only.  In fact,
such a theory can be formulated without a choice of spin structure, using only what we will call a spin-SU(2) structure, which is the SU(2) analog of a $\spinc$ structure.    (For discussion of such
structures, see for example \cite{Bala,FH,CD}.)

The new anomaly
only arises when the theory is formulated in this generality.    If one formulates an SU(2) gauge theory with fermions only on spin manifolds, one will see only the usual anomaly associated
with $\pi_4(\SU)=\Z_2$.   Thus there is a fundamental difference between an SU(2) gauge theory that has the ``old'' anomaly and one that only has the ``new'' anomaly.
An SU(2) gauge theory with the old anomaly -- for example a theory with a single fermion multiplet of isospin 1/2 -- is simply inconsistent, even locally, and it does not make sense to ask about
its dynamics.   An SU(2) gauge theory with only the new anomaly -- for example a theory with an odd number of  fermion multiplets of isospin 3/2 -- is perfectly consistent as a conventional quantum field
theory in Minkowski spacetime, so it makes sense to analyze its dynamics (see \cite{UP} for an attempt).  The anomaly means that this theory cannot be formulated consistently in a general spacetime without some
additional structure and cannot arise as an effective low energy description of  a theory  that microscopically has bosons only.   Examples of the additional structure that do make it possible
to formulate this theory consistently are a choice of a spin or $\spinc$ structure.

Roughly, a theory with the new anomaly is invariant under diffeomorphisms plus gauge transformations that preserve a  spin structure (or some other suitable structure such as a $\spinc$ structure)
  but not under the larger class of diffeomorphisms plus
gauge transformations that preserve a spin-SU(2) structure.

In section \ref{Two}, we review the familiar SU(2) anomaly and describe the new one.    In section \ref{Three}, we describe the effects of gauge symmetry breaking.   A theory with the new SU(2)
anomaly can be Higgsed to a U(1) gauge theory that has a similar anomaly.   This is a theory of ``all-fermion electrodynamics''  \cite{WPS} in which electrons, monopoles, and dyons of odd charges
are all fermions.   Further Higgsing leads to a theory that locally is a $\Z_2$ gauge theory but more globally can be described as a theory with a dynamical spin structure.
In section \ref{Four}, we take a different perspective.   A theory with the new SU(2) anomaly can arise as a boundary state of a five-dimensional theory gapped theory with SU(2) global symmetry.
More interesting and more suprising, it can also arise as a boundary state for a five-dimensional theory with no particular symmetry that is gapped but topologically
non-trivial; its partition function on a closed five-manifold $Y$ is\footnote{Here $w_k(TY)$, $k\in \Z$, which are often denoted 
simply as  $w_k$,  are Stiefel-Whitney classes of the tangent bundle $TY$ of $Y$.}
 $(-1)^{\int_Y w_2(TY) w_3(TY)}$.   The anomalous SU(2) gauge theory gives a novel boundary state for that theory.
Its Higgsing to U(1) gives a boundary state that is already known \cite{McGreevy}; further Higgsing to $\Z_2$ gives a gapped boundary state with a dynamical spin structure.

In most of this paper, we assume for simplicity that all four-manifolds and five-manifolds are orientable; and we consider only the gauge group SU(2).    
In section \ref{Five}, we consider other gauge groups, including Sp(2N), $\SOS$, and Spin(10).   For Sp(2N), one gets a very close analog of what happens for SU(2).
For $\SOS$, we make contact with a recent analysis \cite{CD}.   For Spin(10), we explain that the Grand Unified Theory with that gauge
group can be formulated without a choice of spin structure.     
A generalization to unorientable
manifolds and a similar anomaly  in five dimensions are also briefly discussed in section \ref{Five}.

\section{The Usual SU(2) Anomaly And A New One}\label{Two}

\subsection{Preliminaries}\label{preliminaries}

The Lorentz group SO(1,3) or more precisely its spin double cover $\Spin(1,3)$ has two different irreducible spinor representations, each of dimension two.   They correspond
to fermions  of left-handed or right-handed helicity.
These representations are complex conjugates of
each other, so a left-handed fermion field $\psi_\alpha,\,\alpha=1,2$ in Lorentz signature is the hermitian adjoint of a right-handed fermion field $\t\psi_{\dot \alpha}$, $\dot\alpha=1,2$.

Now consider an SU(2) gauge theory in four dimensions coupled 
to a single multiplet of fermions of isospin 1/2, that is, to a single SU(2) doublet of left-handed fermions $\psi_{\alpha\, i}$ where $\alpha=1,2$ is a Lorentz index and $i=1,2$
is an SU(2) index, and a corresponding doublet of right-handed fermions $\t\psi_{\dot \alpha\,j}$.    
 In Lorentz signature, since the two spinor representations are complex conjugates,
 $\t\psi_{\dot\alpha\,j}$ are simply the hermitian adjoints of the $\psi_{\alpha\,i}$.   The theory can hence be viewed as a theory of eight hermitian fermi fields (the hermitian
parts and $\i$ times the antihermitian parts of $\psi_{\alpha\, i}$).

As in this example, fermion fields in Lorentz signature always carry a real structure; if a fermion field appears in the Lagrangian, so does its hermitian adjoint.  
In Euclidean signature, nothing like that is true in general; what happens depends on the spacetime dimension and the gauge or global symmetry group assumed.   
But four-dimensional fermions in a pseudoreal representation of the gauge group are real in Euclidean signature.   That is because the two spinor representations
of $\Spin(4)$ are each pseudoreal, as is the isospin 1/2 representation of SU(2).   In general, the tensor product of two pseudoreal representations is real, so fermion fields $\psi_{\alpha\,i}$ or $\t\psi_{\dot\alpha\,i}$ valued
in the isospin 1/2 representation of SU(2) -- or more generally in a pseudoreal representation of any gauge and/or global symmetry -- carry a real structure.   Note that
$\psi_{\alpha\,i}$ and $\t\psi_{\dot\alpha\,j}$ are each separately real in Euclidean signature, while in Lorentz signature they are complex conjugates of each other.

Because the fermion fields $\psi_{\alpha,i}$ and $\t\psi_{\dot\alpha,j}$ carry a natural real structure,  the Dirac operator that acts on them is also naturally real.
Let us spell out in detail what this assertion means.   In doing this, we consider a combined 8 component Dirac operator $\slashed{D}_4$
that acts on the four components of $\psi_{\alpha\,i}$ and the four
components of $\t\psi_{\dot\alpha\,j}$ together.\footnote{$\DD_4$ maps $\psi$ to $\t\psi$ and vice-versa, so in a basis $\begin{pmatrix}\psi\cr\t\psi\end{pmatrix}$, one has
 $\DD_4=\begin{pmatrix} 0 & *\cr * & 0\end{pmatrix}$. $\DD_4$ cannot be block-diagonalized in a real basis.}   This Dirac operator is
\be\label{mumz}\DD_4=\sum_{\mu=1}^4 \gamma^\mu\left(\frac{\partial}{\partial x^\mu}+\sum_{a=1}^3 A_\mu^a t_a\right), \ee
where $A_\mu^a$ is the SU(2) gauge field, $\gamma^\mu$, $\mu=1,\dots,4$ are gamma matrices, 
and $t_a$, $a=1,\dots,3$ are antihermitian generators of the SU(2) Lie algebra.    It is not possible to find $4\times 4$ real
gamma matrices obeying the Clifford algebra $\{\gamma_\mu,\gamma_\nu\}=2\delta_{\mu\nu}$.   Similarly it is not possible to find $2\times 2$ real matrices $t_a$ obeying the SU(2)
commutation relations $[t_a,t_b]=\varepsilon_{abc}t_c$ (where $\varepsilon$ denotes the antisymmetric tensor).   
The nonexistence of such matrices is more or less equivalent to the assertion that the spinor representations of $\Spin(4)$
and the isospin 1/2 representation of SU(2) are pseudoreal, not real.      However, to write a real Dirac equation $\DD_4$ for an eight-component fermion field that combines $\psi_{a\,i}$ and $\t\psi_{\dot a\,j}$, we want not $4\times 4$ real
gamma matrices but $8\times 8$ ones.  There is no difficulty with this, and moreover a set of four real $8\times 8$ gamma matrices $\gamma_\mu$ will commute with a set of three real
$t_a$'s that satisfy the SU(2) commutation relations.  This statement is more or less equivalent to the fact that the tensor product of the two representations is real.  

We will also be interested in what happens in five dimensions.   In Lorentz signature, the group $\Spin(1,4)$ has a unique spinor representation, which is a pseudoreal representation of dimension 4.
The tensor product of this representation with the isospin 1/2 representation of SU(2) is real, so it is possible in $4+1$ dimensions to have a theory with a single SU(2) doublet of fermions 
$\psi_{\alpha\,i}$, where now $\alpha=1,\cdots, 4$ is a Lorentz spinor index and $i=1,2$ is an SU(2) index.  This field has eight hermitian components, just as in four dimensions.

Upon going to Euclidean signature, we observe that the spinor
representation of $\Spin(5)$ is pseudoreal, so the field $\psi_{\alpha\,i}$ remains real in Euclidean signature.
The Euclidean signature Dirac operator for this field is therefore also real.   Concretely, if $\gamma_1,\cdots,\gamma_4$ are four real $8\times 8$ gamma matrices that commute
with real SU(2) generators $t_a$, then $\gamma_5=\gamma_1\gamma_2\gamma_3\gamma_4$ is a fifth real $8\times 8$ gamma matrix that commutes with the same SU(2) generators.
So once one has a real four-dimensional Dirac operator (\ref{mumz}), there is no additional difficulty to construct a real five-dimensional Dirac operator
\be\label{mummz}\DD_5=\sum_{\mu=1}^5 \gamma^\mu\left(\frac{\partial}{\partial x^\mu}+\sum_{a=1}^3 A_\mu^a t_a\right). \ee

The only property of the isospin 1/2 representation of SU(2) that was important in any of this is that it is pseudoreal.   So all statements carry over immediately to the
representation of isospin $j$, for any half-integer $j$.

Now let us return in four dimensions to the theory of a single multiplet of fermions of isospin 1/2, that is, the theory of a left-handed fermion field $\psi_{\alpha\,i}$ and its hermitian
conjugate.
Such a field cannot have a bare mass.  Because the spinor representation of the Lorentz group and the $j=1/2$ representation of SU(2) are both pseudoreal, Lorentz
and SU(2) symmetry would force a mass term to be $\varepsilon^{\alpha\beta}\varepsilon^{ij}\psi_{\alpha\,i}\psi_{\beta\,j}+{\mathrm{h. c.}}$
This expression vanishes by fermi statistics, as it is antisymmetric in both Lorentz and gauge indices.

The impossibility of a bare mass means that such a fermion can potentially contribute an anomaly, and this is what actually happens, as we discuss
starting in section \ref{review}.   More generally, the
same holds if the isospin 1/2 representation of SU(2) is replaced by any irreducible, pseudoreal representation of an arbitrary gauge and/or global symmetry group $G$.
In such a representation, a $G$-invariant bilinear form is antisymmetric, so a mass term would have to be $\varepsilon^{\alpha\beta}\kappa^{ij}\psi_{\alpha\,i}\psi_{\beta\,j}+\mathrm{h.c.}$,
with $\kappa^{ij}$ antisymmetric.  This again vanishes by fermi statistics.   In this paper (except in section \ref{diffgroups}), 
the pseudoreal representations we will consider will be the isospin $j$ representations of SU(2), with half-integer $j$.
The absence of bare masses again means that such fermion fields can contribute anomalies.

There is no problem in giving a bare mass to a {\it pair} of fermion multiplets in the isospin $j$ representation of SU(2), for half-integer $j$. (If the two multiplets are $\psi$ and $\chi$,
then obviously fermi statistics do not constrain a mass term $\psi_{\alpha\,i}\chi_{\beta\,j}\varepsilon^{\alpha\beta}\kappa^{ij}$.)  More generally, one can give bare masses to
any even number of isospin $j$ multiplets.   So a possible anomaly can only be a mod 2 effect, sensitive to whether the number of multiplets of a given kind is even or odd.

 By contrast, for integer $j$, the spin $j$ representation of SU(2)  is real and a single multiplet of fermions transforming in that representation could have a bare mass, and hence
could not contribute a gauge anomaly.   For this reason, the interesting SU(2) representations for our purposes in this paper are the ones of half-integer $j$.

\subsection{Review Of The Usual SU(2) Anomaly}\label{review}

The following is the simplest illustration of the usual SU(2) anomaly for a single fermion multiplet of isospin 1/2.
  Consider an SU(2) gauge field of instanton number 1 on a four-sphere $M=S^4$. 
Let $n_L$ and $n_R$ be the number of zero-modes of $\psi_L$ and $\psi_R$, and let $n=n_L+n_R$ be the total number of fermion zero-modes.  

The Atiyah-Singer index theorem says in this situation that 
\be\label{weldo}n_L-n_R=1. \ee
We note that $n=n_L+n_R$ is congruent to $n_L-n_R$ mod 2, so
\be\label{eldo}n\cong 1~~{\mathrm{mod}}~2.  \ee

Having an odd number of fermion zero-modes in an instanton field is a kind of anomaly.  It means that the path integral measure, for integration over the fermions
in the instanton background,  is odd under the operator $(-1)^F$
that counts fermions mod 2.   This operator can be viewed as a diffeomorphism -- a trivial diffeomorphism taken to act on fermions as $-1$ -- or as a gauge transformation
by the central element $-1$ of SU(2).  The second interpretation shows that an anomaly in $(-1)^F$ can be interpreted as a breakdown of SU(2) gauge-invariance.

Obviously, having an odd number of zero-modes, each of which is odd under $(-1)^F$, means that the path integral measure for the zero-modes
is odd under $(-1)^F$.   On the other hand, the path integral measure for  the nonzero fermion modes -- that is, for all fermion modes other than zero-modes -- is invariant
under all symmetries of the classical action, including $(-1)^F$.

This statement is fairly obvious in the present instance, because nonzero modes come in pairs\footnote{If this assertion is unfamiliar, the reader may wish to jump ahead
to eqn. (\ref{pongo}).} and a pair of fermion modes is invariant under $(-1)^F$.   Let us, however,
pause to put the point in a more general perspective.    Fermion anomalies
are elementary to analyze when one considers anomalies in symmetries that leave fixed a given bosonic background.   To understand anomalies in a more general situation
requires more sophisticated arguments, as in \cite{Witten} and later papers.  But in the present paper, it will suffice to consider the easy case of symmetries
that leave fixed the bosonic background.
 What makes this case easy is that if $\varphi$ is any global and/or gauge symmetry that leaves fixed a given bosonic background, then $\varphi$-invariance of the action that
 describes fermions propagating in that background  
\be\label{feldo} I=\int_M \d^4x \sqrt g \overline \psi \i \slashed{D}\psi+\cdots \ee
(where the ellipses refer to possible mass terms or additional couplings)
 means that the nonzero fermion modes are paired up in an $\varphi$-invariant way, ensuring invariance of the path integral measure for those modes.
So a possible anomaly can be computed just by asking whether the measure for the fermion zero-modes is $\varphi$-invariant \cite{thooft,Fujikawa}.

Now let us explain how this anomaly is related to a topological invariant in five dimensions, as described in \cite{Witten}.  In general, let $M$ be a four-manifold
with some SU(2) gauge field and some choice of spin structure.  Let $\varphi$ be a symmetry
whose potential anomaly we want to consider.   Here $\varphi$ might be a diffeomorphism, a gauge transformation, a global symmetry, or a combination thereof.  
We use $\varphi$ to construct a five-manifold $Y$ known as the mapping torus of $\varphi$.     The construction is as follows.   Begin with the five-manifold $Y_0=M\times I$, where $I$ is the unit
interval $0\leq t\leq 1$.  If $\varphi$ is a symmetry of the metric $g$ and gauge field $A$ of $M$, we can take the metric on $Y_0$ to be a product, the gauge field on $Y_0$ to be
a pullback from $M$ (meaning that it is independent of $t$ and has no component in the $t$ direction), and the spin structure of $Y_0$ to be a pullback from $M$.  In general,
if $g$ and $A$ are not $\varphi$-invariant, one picks a metric and gauge field on $Y_0$ that interpolate smoothly between $g,A$ at $t=0$ and $g^\varphi, A^\varphi$ at $t=1$ (where $g^\varphi$ and $A^\varphi$
are the transforms of $g$ and $A$ by $\varphi$).  Finally, to construct $Y$ with its gauge bundle, we glue together the two ends $M\times 0$ and $M\times 1$ of $Y_0$, applying
the symmetry $\varphi$ in the process: thus for $x\in M$, the point $x\times 0$ is identified with $\varphi x\times 1$,   and similarly any bosonic or fermionic field $\chi(x\times 0)$
is set equal to $\chi^\varphi(\varphi x\times 1)$ (where $\chi^\varphi$ is the transform of $\chi$ by $\varphi$).  

The main result of \cite{Witten} is that the fermion path integral for a fermion field on $M$ in a given representation of SU(2) is $\varphi $-invariant if and only if the mod 2
index of the Dirac operator on $Y$ -- in that representation -- vanishes.   Here the mod 2 index is defined as the number of  zero modes of the five-dimensional Dirac operator,
reduced mod 2.    The mod 2 index is governed by a more subtle variant of the usual Atiyah-Singer index theorem \cite{AS}.  For an introduction to the mod 2 index, see \cite{WittenRecent}, section 3.2.
Briefly, whenever one can write a fermion action $\int \bar\psi (\slashed{D}+\cdots)\psi$, the number of fermion zero-modes is always a topological invariant mod 2 (provided one counts all fermion
modes, not distinguishing fermions from antifermions in theories in which such a distinction could be made).   That is because fermi statistics means that the fermion kinetic operator $\slashed{D}+\cdots $ is antisymmetric.  The canonical form
of an antisymmetric matrix is
\be\label{pongo}\begin{pmatrix} 0 & -a & &&&& \cr a&0&&&&&\cr &&0&-b&&&&\cr  &&b&0&&\cr &&&&&&\cr &&&&0&& \cr &&&&&0&\cr &&&&&&\ddots  \end{pmatrix}\ee
with nonzero modes that come in pairs and zero-modes that are not necessarily paired.  When background fields are varied, the number of zero-modes can only change when
one of the skew eigenvalues $a$, $b$, $\cdots$ becomes zero or nonzero.   When this happens, the number of zero-modes changes by 2, so the number of zero-modes mod 2
is always a topological invariant.  It is known as the mod 2 index.    The mod 2 index is {\it not} in general the mod 2 reduction of an integer-valued invariant such as an index;
 indeed, there is no integer-valued
index for a Dirac operator in five dimensions.   

The mod 2 index is difficult to calculate in general; the mod 2 version of the index theorem \cite{AS} provides a somewhat abstract description, not what one would usually regard as
a formula.  However, the mod 2 index
is relatively easy to calculate if one considers a mapping torus built using a symmetry $\varphi $ that leaves invariant the classical background.  Let us consider our example with $\varphi =(-1)^F$.  
This symmetry acts trivially as a diffeomorphism and it also acts trivially on the gauge field $A$.   The mapping torus in this case is just an ordinary product $Y=M\times S^1$,
with a gauge field that is a pullback from $M$.  The meaning of the twist by $(-1)^F $ is just that we should use on $Y$ a spin structure in which the fermions are periodic in going around the $S^1$.
Now we can easily compute the mod 2 index.  To get a fermion zero-mode on $M\times S^1$, we have to take a fermion field that has zero momentum in the $S^1$ direction and that
satisfies the four-dimensional Dirac equation on $M$.   So in this case, the fermion zero-modes in five dimensions are the same as the fermion zero-modes in four dimensions,
and the mod 2 index is simply the number $n$ of fermion zero-modes in four dimensions, reduced mod 2.  Thus, the general statement that the fermion anomaly in four dimensions
is given by the mod 2 index in five dimensions reduces in this situation to the more obvious statement that we began with:  there is an anomaly if the number of fermion zero-modes in the original
four-dimensional problem is odd.  

More generally, 
 the mod 2 index in five dimensions can be computed in four-dimensional terms whenever we consider a symmetry $\varphi $ that leaves fixed the bosonic background.   Before
explaining this, we should explain the following.  As described in section \ref{preliminaries}, in Euclidean signature, a  five-dimensional fermion field in the spin 1/2 representation of SU(2) can be given a real structure, because
the spinor representation of $\Spin(5)$ is pseudoreal, as is the spin 1/2 representation of SU(2).   Any diffeomorphism and/or gauge transformation $\varphi $ preserves this real structure, and this
constrains the possible action of $\varphi $.  A  complex eigenvalue $e^{\i\alpha}$, $\alpha\not=0,\pi$ of $\varphi $ will always be paired with a complex conjugate eigenvalue $e^{-\i\alpha}$, with the complex
conjugate wavefunction.  In addition, $\varphi $ might have eigenalues $\pm 1$, and these are not necessarily paired.   The anomaly of the fermion measure under $\varphi $ is just the determinant of $\varphi $,
regarded as a linear operator on the space of fermion zero-modes.   Since a complex conjugate pair of eigenvalues $e^{\pm \i\alpha}$ does not contribute to this determinant, the determinant
and therefore the anomaly is just $(-1)^{m_{-1}}$, where $m_{-1}$ is the dimension of the space of fermion zero-modes with $\varphi =-1$.  As a check on this, suppose we consider the symmetry
$(-1)^F \varphi $ instead of $\varphi $.  Now we are interested in eigenvalues of $(-1)^F\varphi $ with eigenvalue $-1$.   These are the same as eigenvalues of $\varphi $ with eigenvalue $+1$,
so now the anomaly is $(-1)^{m_1}$, where $m_1$ is the dimension of the space of fermion zero-modes with $\varphi =+1$.  Notice that the total number $n$ of fermion zero-modes is
\be\label{oppo}n\cong m_1+m_{-1}~~~{\mathrm{mod}} ~ 2, \ee because the paired modes with eigenvalues $e^{\pm\i\alpha}$ do not contribute to $n$ mod 2.
So
\be\label{zoppo} (-1)^{m_1}=(-1)^n (-1)^{m_{-1}},\ee
saying, as one would expect, that the anomaly $(-1)^{m_1}$  of the fermion path integral under $\varphi (-1)^F$ is the product of the anomaly $(-1)^n$ under $(-1)^F$ and the anomaly $(-1)^{m_{-1}}$
under $\varphi $.
In particular, if there is no anomaly under $(-1)^F$, then $n$ is even, $m_1\cong m_{-1}$ mod 2, and it does not matter whether we consider the anomaly under $\varphi $ or under $(-1)^F\varphi $.  

If $\varphi $ acts nontrivially as a diffeomorphism (but is still a symmetry of the bosonic background), then the
 mapping torus is not a simple product $M\times S^1$ but a sort of ``semi-direct product'' $M\rtimes S^1$.   The spin structure in the $S^1$ direction
depends on whether the symmetry we want to consider is $\varphi $ or $(-1)^F\varphi $.   If the spin structure in the $S^1$ direction is periodic, then a fermion zero-mode on $M\rtimes S^1$
corresponds to a $\varphi $-invariant zero-mode on $M$, so the five-dimensional mod 2 index is $m_1$.   If the spin structure is antiperiodic, a fermion zero-mode on $M\rtimes S^1$ corresponds to a mode
on $M$ with eigenvalue $-1$ of $\varphi $, so the mod 2 index is $m_{-1}$.  Let us for simplicity consider the case that there is no anomaly in $(-1)^F$, so that $m_1$ and $m_{-1}$ are
equal.  Then
$(-1)^{m_1}=(-1)^{m_{-1}}$, and as we have seen, each of these corresponds to the  transformation under $\varphi $ of the fermion path integral measure in four dimensions.    This confirms,
in this situation, that the mod 2 index in five dimensions governs the anomaly in four dimensions.   (To extend this to the case that there is an anomaly under $(-1)^F$ requires some care
to decide which is which of $\varphi $ and $(-1)^F\varphi $; one has an anomaly and one does not.)

For our purposes in the present paper, it will suffice to consider anomalies under symmetries $\varphi $ that leave fixed a bosonic background, so that the anomaly can be computed
 directly in terms of fermion zero-modes in four dimensions.  This will suffice because every anomaly of interest can be illustrated by a special case with that property.
  The general relation between the anomaly in four dimensions and the mod 2 index in five dimensions can be deduced
by an argument using spectral flow of a family of four-dimensional Dirac operators \cite{Witten}.

\subsection{Generalization To Other Representations}\label{genother}

Now let us consider the generalization of the familiar SU(2) anomaly that we have just reviewed to the case of fermions in some other representation of SU(2).

As we have seen, there is an anomaly when the number of fermion zero-modes in an instanton field is odd.  Moreover, the number of zero-modes mod 2 is given
by the Atiyah-Singer index theorem.  The contribution to the index that depends on the gauge fields is a multiple of $\int_M \Tr\,F\wedge F$, where $F$ is the Yang-Mills
curvature and the trace is taken in the fermion representation.   In a representation of isospin $j$, the quantity $F\wedge F$ is a multiple of $j(j+1)$ and its trace is a multiple of
$j(j+1)(2j+1)$.   The index is 1 for $j=1/2$, so in general it is $\frac{2}{3}j(j+1)(2j+1)$.    The number $n_j$ of zero-modes of a fermion multiplet of isospin $j$ in a gauge field on $M=S^4$
of instanton number 1 is congruent to this mod 2:
\be\label{polno} n_j\cong \frac{2}{3}j(j+1)(2j+1)~~~{\mathrm{mod}}~2. \ee

This number is odd if $j$ is of the form $2r+1/2$, $r\in\Z$, and otherwise it is even.   So the general statement of the usual SU(2) anomaly is that the total number
of fermion multiplets of isospin $2r+1/2$ must be even.   It is possible to have, for example, a consistent theory with a single fermion multiplet of isospin 1/2 and another of isospin 5/2.

The usual SU(2) anomaly does not receive a contribution from a fermion multiplet of isospin $2r+3/2$.   But we will see that there is a more subtle anomaly for a fermion
multiplet of isospin $4r+3/2$, $r\in \Z$.   (The case $4r+7/2$ remains anomaly-free.)

\subsection{Spin-SU(2) Structure}\label{spinsu}

If we consider SU(2) gauge theory on a spin manifold, there is no further anomaly beyond what we have described.  But there is something more subtle that we can do.

Consider an SU(2) gauge theory in which fermions are in half-integer spin representations of SU(2) and bosons are in integer spin representations.   In this case, we can have the option
to formulate the theory on a four-manifold $M$ without a choice of spin structure, choosing a weaker structure that we will call a $\spinSU$ structure.\footnote{Such structures
have been discussed in \cite{Bala,FH,CD}. They are sometimes called $\Spin_h$ structures.}    A $\spinSU$ structure is the
SU(2) analog of a $\spinc$ structure in the case of U(1).   A $\spinc$ structure is appropriate for a theory that, locally, has gauge group U(1) but with the property that fermions have
odd U(1) charge and bosons have even U(1) charge.   This means (in four spacetime dimensions) that not all representations of $\Spin(4)\times \U(1)$ occur in the theory, but only
those representations that are actually representations of 
\be\label{turvo} \Spin_c(4)=(\Spin(4)\times \U(1))/\Z_2. \ee
Here $\Z_2$ is embedded in $\Spin(4)\times \U(1)$ as the product of the element $(-1)^F\in \Spin(4)$ and the element $-1\in \U(1)$.  A $\spinc$ connection is a connection with
structure group $\Spin_c(4)$ and with the property that if one forgets U(1) and projects $\Spin_c(4)=(\Spin(4)\times \U(1))/\Z_2$ to $\Spin(4)/\Z_2=\mathrm{SO}(4)$, the SO(4)
connection is the Riemannian connection for some Riemannian metric on $M$.

A theory in which all fields provide representations of $\Spin_c(4)$ (as opposed to more general representations of $\Spin(4)\times \U(1)$)  can be formulated on a four-manifold
$M$ that is endowed with a $\spinc$ structure (as opposed to a spin structure and a U(1) gauge field).   Concretely, a Riemannian metric plus a $\spinc$ structure give the information
one needs to define parallel transport for fermions of odd charge, and for bosons of even charge.  One does not have the additional structure that would be needed to define parallel
transport for a fermion of even charge (for example, a neutral fermion) or a boson of odd charge.   

Locally, a $\spinc$ connection is equivalent to a U(1) gauge field $A$, but globally this is not the case.   The difference is most striking if $M$ is such as to not admit a spin structure.
The obstruction to a spin structure is the Stiefel-Whitney class $w_2(TM)$, where $TM$ is the tangent bundle of $M$.   Suppose that $S\subset M$ is a closed oriented two-manifold
with $\int_S w_2(TM)\not=0$.  This is an element of $\Z_2$, because $w_2$ is a $\Z_2$-valued cohomology class.   Then the magnetic flux $F=\d A$  for a $\spinc$ connection does not obey conventional
Dirac quantization, but rather
\be\label{welgo}\int_S\frac{F}{2\pi}=\frac{1}{2}\int_S w_2(TM)~~{\mathrm{mod}}~\Z. \ee
This implies that 
 $2F$ does obey  conventional Dirac quantization; $2F$ is the curvature of the gauge connection $2A$, which is a conventional U(1) gauge field that can couple of a charge 2 boson.  
Such a boson is a section of a complex line bundle $\h \L \to M$, and $2A$ is a connection on this line bundle.

A simple example of a four-manifold $M$ with $w_2(TM)\not=0$ is $\CP^2$, which can be parametrized by homogeneous complex coordinates $z_1,z_2,z_3$, obeying $\sum_i |z_i|^2=1$
and with the equivalence relation $z_i\cong e^{\i\alpha}z_i$ for real $\alpha$.   A $\CP^1$ subspace of $\CP^2$ is defined by setting, for example, $z_3=0$.  One has
\be\label{elgo}\int_{\CP^1}w_2(TM)=1~~~\mod ~2.\ee
So to define a $\spinc$ structure, we need 
\be\label{telgo}\int_{\CP^1}\frac{F}{2\pi}=\frac{1}{2}+n,~~~~n\in\Z. \ee
  If we ask for a $\spinc$ structure to be invariant under the SU(3) (actually SU(3)/$\Z_3$ symmetries) of $\CP^2$,
then it is uniquely determined by the flux condition (\ref{telgo}).  
The basic case is to set $n=0$, that is, one-half a unit of flux.   In this case, the complex line bundle $\h\L\to M$ is the fundamental line bundle over $\CP^2$; a charge 2 boson couples
to a single flux quantum on $\CP^1$.   The Atiyah-Singer index theorem can be used to compute the index $\I$ of the Dirac operator for a fermion coupled to a $\spinc$ structure
with flux $f=n+1/2$: 
\be\label{pelgo}\I=\frac{4f^2-1}{8}=\frac{n(n+1)}{2}.\ee
So for example,  for $f=1/2$, one has $\I=0$; for $f=3/2$, one has $\I=1$.    It can be shown, using the positive curvature of $\CP^2$, or by relating the question to algebraic geometry, that this Dirac
equation has zero-modes only of one chirality, so that $\I$ is the total number of  zero-modes for this Dirac quation.

Now let us return to the SU(2) case.   Here we consider a theory in which all fields are representations of what we will call
\be\label{urvo}\Spin_\SU (4)=(\Spin(4)\times \SU)/\Z_2,\ee
rather than more general representations of $\Spin(4)\times \SU$.  By a $\spinSU$ structure, we mean a connection with structure group $\Spin_\SU(4)$, which lets one define
parallel transport for representations of this group, but not for general representations of $\Spin(4)\times \SU$.  (Again the SO(4) part comes from the  Riemannian geometry.)
Any example of a $\spinc$ structure, say one with flux $f$, gives an example of a $\spinSU$ structure, simply by embedding U(1) in SU(2). In other words, if $A$ is any $\spinc$ connnection, then we view
\be\label{toko} \widehat A=\begin{pmatrix} A& 0\cr 0  & -A\end{pmatrix} \ee
as the SU(2) part of a $\spinSU$ connection.   For example, the spin 1/2 representation of SU(2) decomposes under U(1) as the direct sum of representations of charges 1 and charge $-1$, so a fermion
of isospin 1/2 has two components, which respectively see background $\spinc$ connections with flux  $f$ and $-f$, where $f$ is the  flux of $A$.   For another example, the spin 3/2 representation of SU(2) decomposes
under U(1) as the direct sum of representations of charges 3, 1, $-1$, and $-3$, so a fermion of isospin 3/2 has four components, which respectively see $\spinc$ connections of fluxes $3f$, $f$,  $-f$,
and $-3f$.

Using (\ref{pelgo}), we can determine the index  $\I_j$ of the Dirac operator for a fermion of spin 1/2 and isospin $j$ from the case that $\h A$ is constructed from
 a $\spinc$ connection $A$ of flux $f$.   For our purposes, 
it will suffice to consider the basic case $f=1/2$.   We can get $\I_j$ by simply summing the abelian answer (\ref{pelgo}) over the various components of an isospin $j$ fermion field.
This gives
\be\label{mondo}
\I_{1/2}=0, ~~~\I_{3/2}=2,   ~~~\I_j=\sum_{f\in\{j,j-1,\cdots,-j\}}\frac{4f^2-1}{8}= \frac{1}{24}(4j^2-1)(2j+3). \ee
Just as in the $\spinc$ case, the zero-modes of such a fermion are all of the same chirality, and the number of such zero-modes is given by $\I_j$.

For $j=k+1/2$, we have
\be\label{rondo}\I_j=\frac{1}{3}k(k+1)(k+2). \ee
$\I_j$ is always even, so an example of this kind will never give an anomaly in $(-1)^F$.  (For even $k$, a more simple example gives such an anomaly, as reviewed in section \ref{review}.)  However,
we will see that a more subtle anomaly receives a contribution whenever $\I_j$ is not divisible by 4.  From eqn. (\ref{rondo}), the condition for this is that $k\cong 1$ mod 4, or in other
words that $j$ is of the form $4r+3/2$, $r\in \Z$.  

\subsection{The New Anomaly}\label{newanomaly}

To find a new anomaly, we consider the diffeomorphism $\varphi$ of  $\CP^2$ that acts by complex conjugation, that is, by $z_i\to \bar z_i$, where $z_i$ are the homogeneous coordinates of $\CP^2$.
This diffeomorphism reverses the orientation of any subspace $\CP^1\subset \CP^2$, so it reverses the sign of the flux $f$ of any $\spinc$ structure.   Since $f$ is a half-integer, it cannot
vanish, and hence there is no $\varphi$-invariant $\spinc$ structure on $\CP^2$.

However, there is no difficulty to find a $\spinSU$ structure that is invariant under $\varphi$ combined with a suitable gauge transformation.   We just take the $\spinc$ structure that is obtained
by embedding U(1) in SU(2) diagonally as in eqn. (\ref{toko}) and we combine $\varphi$ with the gauge transformation
\be\label{gt}W= \begin{pmatrix} 0 & -1 \cr 1 & 0 \end{pmatrix},
\ee
which also reverses the sign of the flux.  
  We write $\h\varphi$ for the combination of the diffeomorphism $\varphi$ with this gauge transformation. 
  
  Starting with any value of the flux of the underlying $\spinc$ connection, this construction gives a $\h\varphi$-invariant $\spinSU$ structure on $\CP^2$.  For our purposes, we will
  consider the basic case that the $\spinc$ connection has flux  $f=1/2$.    We will consider
  a spin 1/2 fermion of half-integral isospin $j$ coupled to this $\spinSU$ structure.   We want to know if there is a quantum anomaly in the action of $\h\varphi$. 
  
 Since $\h\varphi$ is a symmetry of the classical background, all we have to do, as explained in  section \ref{review}, is to compute the determinant of $\h\varphi$ in its action on the space
 of fermion zero-modes.   The following fact will simplify the analysis.  For any half-integer $j$, the space of zero-modes of the Dirac operator acting on a fermion field of spin 1/2
 and isospin $j$ is real.   That is simply because the spin 1/2 representation of $\Spin(4)$ is pseudoreal, and the spin $j$ representations of SU(2) are also pseudoreal, so their
 tensor products are real representations of the group $\Spin_{\SU}(4)=(\Spin(4)\times \SU)/\Z_2$.
 
 The space of zero-modes has some additional crucial properties.   
 It has a U(1) symmetry that comes simply 
from constant diagonal gauge transformations in SU(2).   These are symmetries of the $\spinSU$ structure, which was constructed in eqn. (\ref{toko}) using a diagonal embedding of U(1)
in SU(2).
 Moreover, all zero-modes have odd U(1) charges, and in particular none of them are U(1)-invariant.   That is so simply because the spin $j$ representation of SU(2) decomposes under U(1)
 as a sum of components of charges $2j, 2j-2,\cdots, -2j$, all of which are odd (for half-integer $j$).   
 Finally $\h\varphi$ commutes with the diagonal U(1) generator.   That is because the gauge transformation $W$
 in (\ref{gt}) that is used in defining $\h\varphi$ anticommutes with the generator of the diagonal U(1) symmetry in  eqn. (\ref{toko}).   Of course, the reason for this is that $W$
  was chosen to reverse the sign of the U(1) flux, to compensate for the fact that this flux is odd under $z_i\to \bar z_i$. 
 
 The symmetry $\h\varphi$ acting on the isospin $j$ representation of $(\Spin(4)\times \SU)/\Z_2$ obeys $\h\varphi^2=1$.  This fairly subtle fact can be understood as follows.  As a symmetry
 of $\CP^2$, $\varphi$ has a codimension 2 fixed point set on which the $z_i$ are all real.    Since the fixed point set is of codimension 2, near a fixed point, we can pick real coordinates
 $x_1,\cdots x_4$ such that $x_3,x_4$ are odd under $\varphi$ and $x_1,x_2$ are even.    Accordingly, $\varphi$ acts near the fixed point set by $(x_1,x_2,x_3,x_4)
 \to (x_1,x_2,-x_3,,-x_4)$.  One can think of this as  a $\pi$ rotation of the $x_3-x_4$ plane; its square is a $2\pi$ rotation, and acts on fermions as $-1$.   But $\h\varphi$ is defined
 by combining the geometrical transformation $\varphi$ with the gauge transformation in (\ref{gt}), which also squares to $-1$.  So $\h\varphi^2=+1$.  Note that there is no freedom to modify
 $W$ to change this conclusion.   Since $W$ has to anticommute with a diagonal U(1) generator to make the construction possible, and since we want $\det W=1$ so that $W$ is valued
 in SU(2),   there is no way to avoid the fact that $W^2=-1$.   
 
Since $\h\varphi^2=1$ and $\h\varphi$ anticommutes with the U(1) generator,
 $\h\varphi$ and the unbroken U(1) together generate a group of symmetries of the space of fermion zero-modes that is isomorphic to O(2).   
 The group O(2) has irreducible representations of dimension 1 in which U(1) acts trivially, and irreducible representations of dimension 2 in which U(1) acts nontrivially.
 
 In a two-dimensional real representation of O(2) in which U(1) acts with charges $\pm n$, the U(1) generator is, in a suitable basis
 \be\label{toggo}\begin{pmatrix} 0 & -n\cr n& 0 \end{pmatrix}.\ee
 And $\h\varphi$ is conjugate by an element of U(1) to 
  \be\label{toggon}\begin{pmatrix} 1 & 0\cr 0& -1 \end{pmatrix}.\ee
  In particular, in any such representation of O(2), $\h\varphi$ has one eigenvalue $+1$ and one eigenvalue $-1$; its determinant in this two-dimensional space is $-1$.
A quicker way to explain this is that since $\h\varphi^2=1$, its possible eigenvalues are 1 and $-1$; since $\h\varphi$ anticommutes with the U(1) generator, it has one eigenvalue of each type.  

The dimension of the space of zero-modes is $\I_j$, defined in eqns. (\ref{mondo}) and (\ref{rondo}); since there are no 1-dimensional representations of O(2) among the zero-modes,
the space of zero-modes consists of $\I_j/2$ representations each of which is of dimension 2.    The determinant of $\h\varphi$ in each such 2-dimensional representation is $-1$.
So its determinant in the full space of zero-modes is
\be\label{zoggo}\det\,\h\varphi =(-1)^{\I_j/2}. \ee

When this determinant is $-1$, the path integral measure is odd under the combined diffeomorphism plus gauge transformation $\h\varphi$.    This is our anomaly.
An anomaly-free theory has $\det\,\h\varphi=+1$.    In view of eqn. (\ref{rondo}), in a theory that is free of this anomaly, the total number of fermion multiplets of isospin $4r+3/2$ is even. 
This is in addition to requiring that the total number of fermion multiplets of isospin $2r+1/2$ should be even, to avoid the usual SU(2) anomaly.

\subsection{Five-Dimensional Interpretation Of The New Anomaly}\label{fived}

The new SU(2) anomaly has a five-dimensional interpretation that is quite analogous to what we explained for the old one.

We use the diffeomorphism plus gauge transformation $\h\varphi$ to build a  mapping torus, which in this case we may describe as a 
``semi-direct product''\footnote{This manifold is a special case of the Dold manifold $(\CP^n\times S^m)/\Z_2$, where $\Z_2$ acts as complex comjugation on
$\CP^n$ and as the antipodal map on $S^m$.} $\CP^2\rtimes S^1$, with a certain spin-SU(2) structure.   We recall that this is constructed starting with $\CP^2\times I$, where
$I$ is a unit interval.  The metric of $\CP^2\times I$ is taken to be a product metric
and the $\spinSU$ structure is pulled back from $\CP^2$.   Then one glues together the two ends of $\CP^2\times I$, via the diffeomorphism plus gauge transformation $\h\varphi$,
to make a closed  five-manifold $\CP^2\rtimes S^1$ with a $\spinSU$ structure.  It is fibered over $S^1$ with fiber $\CP^2$; the monodromy when one goes around the $S^1$ is $\h\varphi$.

A fermion zero-mode on $\CP^2\times I$ is a $\h\varphi$-invariant fermion zero-mode on $\CP^2$.     So in view of the analysis in section \ref{newanomaly}, the five-dimensional mod
2 index is $\I/2$, where $\I$ is the dimension of the space of zero-modes in four dimensions.     Thus, the mod 2 index in five dimensions is nonzero precisely when $\I/2$ is odd,
that is, precisely when the theory is anomalous.  

The following is one systematic way to understand the fact that the old and new SU(2) anomalies both  have five-dimensional interpretations in terms of a mod 2 index.   In general, anomalies
for fermions in $D$ dimensions are always governed by a corresponding $\eta$-invariant in $D+1$ dimensions.  This was originally shown in \cite{GGA}, with an important later refinement in \cite{DF}.
For a theory with no perturbative anomaly, the $\eta$-invariant in question is a cobordism invariant, so in particular, for $D$-dimensional fermions with no perturbative anomaly,
a possible global anomaly is governed by a cobordism invariant in $D+1$ dimensions.   When the relevant $D+1$-dimensional Dirac operator is real -- and in certain other situations --
the $\eta$-invariant reduces to a mod 2 index.   (This is explained in \cite{WittenRecent}, for example.)   We are in that situation for the old and new SU(2) anomalies in four dimensions.

We found the new SU(2) anomaly as an anomaly in a combined gauge transformation plus diffeomorphism in a particular example.   The new SU(2) anomaly is never an anomaly in a gauge 
transformation only.   A partial explanation of this is that to test for an anomaly under a gauge transformation $\varphi$  of some $\spinSU$ bundle on a four-manifold $M$, we would have
to compute a mod 2 index on $M\times S^1$, with a $\spinSU$ structure that depends on $\varphi$.  But in section \ref{bsu}, by relating the new anomaly to the five-dimensional invariant
$\int w_2w_3$, we will show that it always vanishes on $M\times S^1$, regardless of the $\spinSU$ structure.   For a more complete explanation, we show in section \ref{notenew} how to define the partition
function of a theory with the new SU(2) anomaly on a four-manifold $M$ endowed with suitable additional structure (a spin or $\spinc$ structure).    This shows that the theory is invariant under
SU(2) gauge transformations, though not in general under diffeomorphisms that do not preserve the additional structure used in defining it.

\subsection{No More Such Anomalies}\label{nomore}

We have identified two types of anomaly in SU(2) gauge theory in four dimensions.   The simplest example of the usual SU(2) anomaly is a theory with a single
multiplet of fermions in the isospin 1/2 representation of SU(2).   The simplest example of the new anomaly is a theory with a single multiplet of fermions in the isospin 3/2 representation of SU(2).

In each case these are discrete, mod 2 anomalies.   A mod 2 anomaly in four dimensions should be related to a mod 2 topological invariant in five dimensions.   We have identified the relevant
invariants.    Let $j$ be a half-integer\footnote{This is the interesting case, as explained at the beginning of section \ref{review}.  For integer $j$, one can still formally define a mod 2 index of
the five-dimensional Dirac operator,
but it vanishes identically.   This is clear from the fact that a bare mass is possible for such a field.   A direct proof can be given by showing that the Euclidean Dirac operator for such
a field has an antilinear symmetry whose square is $-1$, so by a Euclidean analog of Kramers doubling, the space of its zero-modes has even dimension.   See \cite{WittenRecent} for
such arguments.} 
and denote as $\ii_j$ the mod 2 index in five dimensions for a fermion in the isospin $j$ representation.
   The five-dimensional invariants associated to the usual SU(2) anomaly and the new one are
$\ii_{1/2}$ and $\ii_{3/2}$.

We have shown that $\ii_{1/2}$ and $\ii_{3/2}$ are different from each other and are each nontrivial by giving examples.   An example with $\ii_{1/2}\not=0$, $\ii_{3/2}=0$ is provided
by $Y_1=S^4\times S^1$, with an SU(2) bundle of instanton number 1 on $S^4$.   An example with $\ii_{1/2}=0$, $\ii_{3/2}\not=0$ is provided by $Y_2=\CP^2\rtimes S^1$, as discussed
in section \ref{fived}.

What about $\ii_j$ with $j>3/2$?   Are these new invariants?   The answer to this question is ``no.''    In any dimension, the mod 2 index of the Dirac operator is a cobordism invariant whenever
it can be defined.   Thus in particular for any half-integer $j$,  $\ii_j$ is a cobordism invariant of a five-manifold $Y$ with $\spinSU$ structure.   (This statement means that if $Y$ is the boundary
of a six-manifold $Z$ and the $\spinSU$ structure of $Y$ extends over $Z$, then $\ii_j(Y)=0$ for all $j$.)    On the other hand, as shown in \cite{GPW,WW}, slightly extending
computations in \cite{FH}, the group of cobordism invariants
for five-manifolds with $\spinSU$ structure is $\Z_2\times \Z_2$.  So any set of two independent $\Z_2$-valued cobordism invariants for such manifolds is a complete set.   In particular, $\ii_{1/2}$ and
$\ii_{3/2}$ are a complete set of invariants, and any $\ii_j$ can be expressed in terms of them.   In view of the computations in section \ref{Two} for the special cases of $Y_1$ and $Y_2$,
we have in general  $\ii_{2r+1/2}=\ii_{1/2}$, $\ii_{4r+3/2}=\ii_{3/2}$, and $\ii_{4r+7/2}=0$.   

Among our examples, $Y_1$ is a spin manifold but $Y_2$ has only a $\spinSU$ structure.    Thus our examples show that $\ii_{1/2}$ can be nonzero on a spin manifold but do
not demonstrate this for $\ii_{3/2}$.   In fact, $\ii_{3/2}$ is identically zero on a spin manifold.  This follows from the fact  that the group of cobordism invariants for a five-dimensional
spin manifold endowed with an SU(2) bundle is $\Z_2$ \cite{FH,GPW,WW}.   $\ii_{1/2}$ is a non-trivial invariant for this cobordism problem, so any other invariant is either trivial or is equal to $\ii_{1/2}$.
To decide which is the case, it suffices to evaluate the invariant for $Y_1$.  Since $\ii_{3/2}$ vanishes for  $Y_1$, it is identically zero for spin manifolds.     For a manifold that admits
only a $\spinSU$ structure and not a spin structure, $\ii_{3/2}$ can be nonzero; in section \ref{bsu} we will show that on such a manifold, $\ii_{3/2}$ is actually independent of the choice of $\spinSU$
structure.

\section{Gauge Symmetry Breaking }\label{Three}

\subsection{Gauge Symmetry Breaking To U(1)}\label{gsbone}

Now let us consider an SU(2) gauge theory which is consistent -- because it lacks the original SU(2) anomaly -- but is affected by the more subtle anomaly that we found in section \ref{newanomaly}.
Such a theory is consistent if formulated on a spin manifold, so it can be studied sensibly.    But it cannot be consistently formulated on a manifold with only a $\spinSU$ structure.  
A simple example is an SU(2) gauge theory with a single fermion multiplet  of isospin 3/2.  

We can introduce a Higgs field $\vec\phi$  in the spin 1 representation of SU(2).   Its expectation value can spontaneously break SU(2) to U(1).    The resulting U(1) theory must somehow
embody the same anomaly:  it must be consistent if formulated on a spin manifold, but not if formulated on on a bare four-manifold with no additional structure.    We would like to see how this comes about.

Symmetry breaking from SU(2) to U(1) changes the nature of the problem, as follows.   A single isospin 3/2 fermion multiplet cannot receive an SU(2)-invariant bare mass.   Hence
it is possible for the path integral of such a field to be anomalous, and this is what we have found in section \ref{newanomaly}.  However, once SU(2) is spontaneously broken to U(1),
the isospin 3/2 fermion can gain a bare mass by coupling to $\vec \phi$.  Hence we can integrate out the fermions, and it must be possible to see the anomaly in terms of effective
couplings of a U(1) gauge theory only.   In fact, what is relevant here is an anomalous theory that is known as ``all-fermion electrodynamics.''   This anomaly was first demonstrated in \cite{WPS};
we will give an explanation of it in section \ref{allfer}.

The low energy theory that we will get after symmetry breaking from SU(2) to U(1) has  a full set of electric and magnetic
charges allowed by Dirac quantization; the 't Hooft-Polyakov monopole of the theory has the minimum possible magnetic charge relative to the component of the elementary fermion field
of electric charge 1.

In general, a U(1) gauge theory might have fermions that carry neither electric nor magnetic charge.   But that is not the case here; because we are considering a theory in which
all elementary fields are in representations of $(\Spin(4)\times \SU)/\Z_2$, all magnetically neutral states (at least in weakly coupled perturbation theory) are fermions of odd electric
charge, or bosons of even electric charge.   

 In general, in a theory that does not have any neutral fermions, one can ask whether an electrically charged particle with electric and magnetic charges $(1,0)$ is a boson
or a fermion.  (The answer will be the same for each such particle, since otherwise a particle-antiparticle pair would be a neutral fermion.)
Likewise one can ask  the same question for a monopole with charges $(0,1)$ or a dyon with charges $(1,1)$.   Duality in the low energy effective U(1) gauge theory permutes these three particle
types
so any general constraints are invariant under such permutations.

If  two of these basic particles are bosons or two are fermions, then after taking into account the angular momentum in the electromagnetic field, we can deduce that the third is a fermion;
likewise if one is a boson and another is a fermion, then the third is a boson.   Thus on this grounds alone, there are two possibilities: two of the three basic charges are bosons
and one is a fermion, or all three are fermions.    The first case is anomaly-free, as we will discuss in a moment, but the second case, which is ``all-fermion electrodynamics,'' is anomalous.   

To show that the first case is anomaly-free, it suffices to give a manifestly anomaly-free realization.  For this, we simply consider an SU(2) gauge theory with the Higgs triplet field
$\vec\phi$ that breaks SU(2) to U(1), and add two multiplets of spin 1/2 fermions in the isospin 1/2 representation of the gauge group.  This is actually a much-studied
theory.     It is anomaly-free (and fermion bare masses are possible), because we included an even number of identical fermion multiplets.   The elementary fermions have electric and magnetic
charges $(\pm 1,0)$ and are, of course, fermions.   The  't Hooft-Polyakov monopole is rotationally invariant at the classical level (up to a gauge transformation), so in the absence of
fermions, it is a boson.   Fermion zero-modes can give the monopole
unusual quantum numbers  under global symmetries\footnote{Concretely \cite{JR}, after giving the fermions a mass
by coupling to $\vec\phi$, one can be left with an SO(2) global symmetry that rotates them.   Monopoles can have half-integral charge under this SO(2).}  but  the zero-modes of isospin 1/2
fermions do not carry angular momentum and the monopole is a spinless boson.   The (1,1) dyon is therefore also a boson.  Consistency of this theory makes
clear that in general a U(1) gauge theory in which two of the basic charges are bosons and one is a fermion can be consistent.

However, we will now see that the anomalous SU(2) gauge theory with a single fermion multiplet of isospin 3/2 reduces, after symmetry breaking, to the all-fermion case, which is anomalous.
Since the particles in this theory with minimum electric charge are elementary fermions, we just have to show that the 't Hooft-Polyakov monopole is also a fermion.   The Callias index theorem
\cite{Callias} implies that a single isospin 3/2 fermion multiplet in the field of an 't Hooft-Polyakov monopole of magnetic charge 1 has a four-dimensional space of zero-modes.\footnote{This assertion,
and its generalization for any isospin $j$,  follows from the formula for ${\mathrm{index}}(L)$ on p. 231 of \cite{Callias}.   In that formula, one has to take $T=j$.
 One also takes $m=0$ (because we are interested in a case in which a bare mass $m$ is not possible) and hence $\{m\}=-1/2$.}     The 't Hooft-Polyakov
monopole is spherically symmetric (up to a gauge transformation), so the four zero-modes furnish a representation of the rotation group.
  Allowing for the angular momentum in the electromagnetic field, an isospin 3/2 fermion in the field of the monopole
carries integer angular momentum.  A four-dimensional representation of the rotation group  with only integer spin appearing is  the sum of either  four copies of  spin 0 or one copy of spin 0
and one copy of spin 1.   As we will explain in a moment, in this particular case, the four zero-modes transform as spin $0\oplus 1$.

Having described the zero-modes, 
we follow the logic of \cite{JR} for quantization of the monopole.   Quantization turns the classical zero-modes into quantum operators that obey the anticommutation relations
of Dirac gamma matrices.   Thus in the present case, we have a single gamma matrix $\gamma_0$ that commutes with the rotation group and a triplet of gamma matrices $\vec\gamma$ of spin 1.
To represent $\vec\gamma$, we need a two-dimensional Hilbert space that transforms as spin 1/2 under rotations; the operators $\vec\gamma$ act in this space like the Pauli angular momentum
matrices $\vec\sigma$.  The fourth gamma matrix $\gamma_0$, because it anticommutes with $\vec\gamma$ but carries no angular momentum, forces a doubling of the spectrum
but without changing the fact that the states have spin 1/2.    Thus the 't Hooft-Polyakov monopole in this theory has angular momentum 1/2 and is a fermion.   Accordingly, the dyon
is also a fermion and this theory gives a realization of all-fermion electrodynamics.

The statement that the four zero-modes have spin $0\oplus 1$ rather than $0\oplus 0\oplus 0\oplus 0$ can be deduced from sections 3.1 and 3.2 of \cite{MRB}.   A less technical explanation is as follows.
A spin 1/2 fermion has components of helicity $\pm 1/2$.  In the absence of the monopole, the minimum angular momentum of a particle of helicity $h$ is $|h|$.   However, if the particle
has electric charge $q$ and interacts with a monopole of charge $g$, then its minimum angular momentum is shifted from $|h|$ 
 to $|h+qg/4\pi|$.   To get angular momentum zero, we need $h+qg/4\pi=0$.   For an elementary fermion
field that transforms under the SU(2) gauge symmetry with isospin 3/2, the possible values of $q$ are $3,1,-1,-3$, and the corresponding values 
 of $qg/4\pi$ (in the field of a minimum charge monopole) 
 are $3/2,1/2,-1/2, $ and $-3/2$.  To get $h+qg/4\pi=0$, we take $h=1/2$, $q=-1$, or $h=-1/2$, $q=1$.  Thus in the partial wave decomposition of the elementary fermion field of this
 theory, angular momentum 0 occurs precisely twice.  Since the Dirac equation is a first order differential equation, a given partial wave can lead to at most 1 zero-mode,
 and actually since only one llnear combination of the two partial waves of spin 0 obeys the boundary condition at the origin, those two partial waves together cannot possibly give more than 1 zero-mode.
 So there is no way to make four zero-modes of spin 0, and the zero-modes in this problem carry angular momentum $0\oplus 1$, as assumed in the derivation of all-fermion electrodynamics.
 
 More generally, consider an elementary fermion multiplet that couples to the SU(2) gauge field with any half-integral isospin $j$.   The Callias index theorem says that in the field of a monopole,
 such a fermion has $(j+1/2)^2$ zero-modes.  These modes transform under rotations with spin
 $0\oplus 1\oplus \cdots \oplus (j-1/2)$.  Quantizing these $(j+1/2)^2$ zero-modes gives states of half-integer angular momentum if and only if $j$ is of the form $4r+3/2$, $r\in \Z$.  
 If and only if the total number of fermion multiplets of isospin $4r+3/2$ is odd, the 't Hooft-Polyakov monopole turns out to be a fermion and we get a model of all-fermion electrodynamics.
 
 The minimal example that illustrates the new anomaly is an SU(2) gauge theory with a single multiplet of fermions of isospin 3/2.   This theory is asymptotically free (just barely, that is the
 coefficient of the one-loop beta function is just slightly negative), so it gives an ultraviolet completion of all-fermion electrodynamics.

 \subsection{The Anomaly In All-Fermion Electrodynamics}\label{allfer}
 
 Here we will give a direct explanation of the anomaly in all-fermion electrodynamics.\footnote{This anomaly has been discussed in \cite{T,OneForm}.  The fact
 that the interaction (\ref{rolb}) makes the monopole a fermion is explained in \cite{T} and in   \cite{CD},  section 4.3.}
 
 We work on a four-manifold $M$ without a chosen spin structure.    The choice of a spin structure would trivialize the discussion of all-fermion electrodynamics (just as it trivializes the new SU(2)
 anomaly discussed in the present paper) because it means that a neutral fermion is possible and therefore that a quasiparticle of any given electric and magnetic charges can be either a boson
 or a fermion.  (Less obviously, as explained at the end of the present section, the choice of a background $\spinc$ structure on $M$ also trivializes the discussion.)
 
 The most obvious form of electrodynamics is ordinary U(1) gauge theory of an ordinary U(1) gauge field.    In the absence of a spin structure, a particle with electric
 charge but no magnetic charge then has to be a boson.
 
 We already discussed in section \ref{spinsu} how to modify this if we want a field of electric charge 1 to be fermionic rather than bosonic: the gauge field must be a $\spinc$ connection
 rather than an ordinary U(1) gauge field.
 
 Let us, however, go back to ordinary U(1) gauge theory and ask what one has to do in order to make a charge 1 magnetic monopole a fermion.
 The answer is that this will happen if the path integral measure of the low energy theory has a factor that roughly speaking is
 \be\label{dolbo} (-1)^{\int_M w_2(TM) \cdot F/2\pi}. \ee
 This formula needs some improvement, since $F/2\pi$ is a differential form and $w_2(TM)$ is a $\Z_2$-valued cohomology class.   A more accurate description is as follows.
 The gauge field $A$ is a connection on a complex line bundle $\L$ over spacetime.   This line bundle has a first Chern class $c_1(\L)\in H^2(M,\Z)$.  Reducing it mod 2,
 we get a class in $H^2(M,\Z_2)$, which we also denote as $c_1(\L)$.   Then taking a cup product in $\Z_2$-valued cohomology, we can define the mod 2 invariant
 $\int_M w_2(TM)\cdot c_1(\L)$, and the factor that we want in the path integral is \be\label{rolb} (-1)^{\int_M w_2(TM)\cdot  c_1(\L)}. \ee  (Another version of this formula is described
 in section \ref{detail}.)
 To be more exact, (\ref{rolb}) is a topological invariant assuming that $M$ is compact and without boundary.   We make this assumption to simplify the analysis.   Otherwise it is necessary
 to specify the behavior along the boundary of $M$ or at infinity.
 
 Now we want to include a monopole. 
  Since we have assumed that $M$ is a compact four-manifold without boundary, it is natural
 to take the worldline $\ell$ of the monopole to be a circle.
   In the low energy U(1) gauge theory, the worldline  of a charge 1 monopole is represented by a charge 1 
 't Hooft operator.   Concretely, to compute in the presence of this 't Hooft operator,  we do U(1) gauge theory on the complement of $\ell$, but along $\ell$, the U(1) gauge field has a singularity that is characterized by
 \be\label{wolb} \d F=2\pi \delta_\ell, \ee
 where $\delta_\ell$ is a three-form delta function that is Poincar\'e dual to $\ell$.

The line bundle $\L$ and the corresponding cohomology class $c_1(\L)$ are only defined on the complement of $\ell$, not on all of $M$, Therefore, the formula (\ref{rolb}) does not
make sense in the presence of the 't Hooft operator.

To try to remedy the situation, we can remove from $M$ a small tubular  neighborhood of $\ell$, replacing $M$ with a manifold $M'$ with boundary on which $\L$ is defined.
The boundary of $M'$ is a copy of $S^2\times \ell$. The fact that the 't Hooft operator carries magnetic charge 1 means that
\be\label{wolbo}\int_{S^2}\frac{F}{2\pi}=1. \ee

 We now replace (\ref{rolb}) with
\be\label{otolb} (-1)^{\int_{M'} w_2(TM')\cdot  c_1(\L)}. \ee
We have made some progress, because $w_2(TM)$ and $c_1(\L)$ both make sense as elements of $H^2(M',\Z_2)$.   However, we still have a problem, because $M'$ is a manifold
with boundary.   To integrate a cohomology class -- in this case $w_2(TM)\cdot c_1(\L)$ -- on a manifold with boundary, we need a trivialization of the cohomology class on the boundary.

In the present case, since $w_2(TM')\cdot c_1(\L)$ is the product of two factors, we could {\it a priori} trivialize it along $\partial M'=S^2\times \ell$ by trivializing either of the two factors.
However, we cannot trivialize $c_1(\L)$ along $\partial M'=S^2\times \ell$, because eqn. (\ref{wolbo}) says that the restriction of $c_1(\L)$ to $\partial M'$ is nontrivial.  The remaining
option is to trivialize $w_2(TM')$ along $\partial M'$.  

A trivialization of $w_2(TM')$ is a spin structure.  We have learned that to define the factor (\ref{otolb}), we need a spin structure along $\partial M'=S^2\times \ell$.  Since $S^2$ is simply
connected, there are precisely two spin structures along $S^2\times \ell$ -- differing by the sign that arises in parallel transport of a fermion all the way around $\ell$.  Since the $S^2$ plays no role
in classifying spin structures on $\partial M'=S^2\times \ell$, the choices of a a spin structure along $S^2\times \ell$ are directly in correspondence with the choices of a spin structure along
$\ell\subset M$ (or equivalently in a tubular neighborhood of $\ell\subset M$).

We have learned that if the integrand of the path integral contains the factor (\ref{dolbo}), then to define the propagation of a charge 1 monopole around the worldline $\ell$,
we need a spin structure along $\ell$.   In other words, the charge 1 monopole is a fermion.   

So to make the elementary electric charge a fermion, we take the gauge field to be a $\spinc$ connection rather than a U(1) gauge field.   To make the elementary monopole a fermion,
we need to include in the integrand of the path integral the factor (\ref{rolb}). 

To get all-fermion electrodynamics, we most do both of these.  But here there is a problem.   If the gauge field $A$ is really a $\spinc$ connection, then there is no ``charge 1'' line bundle $\L$
whose curvature would be $F/2\pi$.   Likewise there is no integral cohomology class $c_1(\L)$.    So in the $\spinc$ case, we cannot define the factor (\ref{rolb}).   If $A$ is a $\spinc$ connection,
then $2A$ is an ordinary U(1) gauge field, and there exists a complex line bundle $\h\L$ with connection $2A$, which would reduce to $\L^2$ in the case of U(1) gauge theory.
One can then define an integral cohomology class $c_1(\h \L)$, which would correspond to $2c_1(\L)$ in U(1) gauge theory.   But in the $\spinc$ case,
 there is no way to eliminate the factor of 2 and thus
no way to define the interaction (\ref{rolb}).   This is a topological explanation of why all-fermion electrodynamics is anomalous, that is, not well-defined.   The anomaly is trivialized by a choice
of spin structure, because such a choice eliminates the distinction between a U(1) gauge field (for which eqn. (\ref{rolb}) can be defined) and a $\spinc$ connection (for which it cannot be).  

The choice of a background, non-dynamical $\spinc$ connection $A_0$ would trivialize the discussion of all-fermion electrodynamics for the following reason.   By subtracting $A_0$ from the
dynamical $\spinc$ connection $A$, we would get an ordinary U(1) gauge field $A'=A-A_0$.    Therefore a charge 1 bosonic quasiparticle is possible.   Once such a quasiparticle is possible,
it follows that a quasiparticle of any electric and magnetic charges can be either a boson or a fermion.

\subsubsection{Some More Detail On The Amplitude For Monopole Propagation}\label{detail}

To complete this explanation, we would like to show that if the spin structure of $\ell$ is changed, this changes the sign of the amplitude for a monopole to propagate around $\ell$. 
We will give two versions of the explanation.   For an abstract explanation, two trivializations $\beta_1$ and $\beta_2$ of a class $w_2\in H^2(M',\Z_2)$ differ by an element $a\in H^1(M,\Z_2)$.
In the present case, the trivializations (or spin structures) $\beta_1$ and $\beta_2$ are defined only near $\partial M'$, so likewise $a$ is defined only near $\partial M'$.    For $i=1,2$, let
$\int_{M';\beta_i} w_2(TM')\cdot c_1(\L)$ be the integral in question defined using the trivialization $\beta_i$ along $\partial M'$.    The general topological formula in this situation is
\be\label{dobso}\int_{M';\beta_1}w_2(TM')\cdot c_1(\L)-\int_{M';\beta_2} w_2(TM')\cdot c_1(\L)=\int_{\partial M'}a\cdot c_1(\L)=\int_{\ell\times S^2}a \cdot c_1(\L). \ee
If the two spin structures along $\partial M'$ are different, this means that $\int_\ell a=1$, and from eqn. (\ref{wolbo}), we have\footnote{All statements are mod 2.}
 $\int_{S^2}c_1(\L)=1$.   So the right hand side of eqn. (\ref{dobso}) is 1.
Hence flipping the spin structure along the monopole worldline $\ell$ shifts what we mean by $\int_M w_2\cdot c_1(\L)$ by 1, and therefore  it changes the sign of the factor  (\ref{rolb})  in the path
integral measure.   This is the expected effect of flipping the spin structure if the monopole is a fermion.  

For a somewhat more concrete version of this explanation, we will first explain the notion of an integral lift of $w_2(TM)$, which will also be useful in section \ref{Four}.
We look at the short exact sequence of abelian groups
\be\label{nunx} 0\to \Z \overset{2}{\rightarrow} \Z\overset{r}{\rightarrow}\Z_2\to 0,\ee
 where the first map is multiplication by 2 and the second is reduction mod 2.
 This leads to a long exact sequence of cohomology groups that reads in part
 \be\label{imx}\cdots \to H^2(M,\Z)\overset{r}{\rightarrow} H^2(M,\Z_2) \overset{\beta}{\rightarrow} H^3(M,\Z)\to\cdots. \ee
 A class $x\in H^2(M,\Z)$ is called an integral lift of $w_2(TM) \in H^2(M,\Z_2)$ if $w_2(TM)=r(x)$.   From the exactness of the sequence (\ref{imx}), such an $x$ exists if and only if
 $W_3(TM)=0$, where
 \be\label{wimx} W_3(TM)=\beta(w_2(TM)). \ee
 The mod 2 reduction of $W_3(TM)$ is the Stiefel-Whitney class $w_3(TM)$.  This fact will be important in section \ref{Four}.
 
 For the moment, however, we are interested in the case that $W_3(TM)=0$.   In fact, this is always true in four dimensions.  Given
  $W_3(TM)=0$, we pick any integral lift $x$ of $w_2(TM)$.   There will
 then be a complex line bundle $\RR\to M$ with $x=c_1(\RR)$.   We write $B$ for a connection on $\RR$, and $G=\d B$ for the corresponding field strength.  The fact that $x$ is congruent mod 2
 to $w_2(TM)$ means that for any closed oriented two-surface $S\subset M$,
 \be\label{zimx} \int_S\frac{G}{2\pi}=\int_S w_2(TM)~~~\mod~2.  \ee
 Comparing to eqn. (\ref{welgo}), we see that $C=\frac{1}{2}B$ can be regarded as a $\spinc$ connection, and thus the existence of a $\spinc$ structure on $M$ is equivalent to the existence
 of an integral lift of $w_2(TM)$.
 
 For now, though, we continue the discussion with the ordinary U(1) gauge field $B$ and its field strength $G$.   Eqn. (\ref{zimx}) means that if $M$ has no boundary (and no monopole singularity)
 we can write
 \be\label{flimx}\int_M w_2(TM)\cdot c_1(\L)=\int_M \frac{G}{2\pi}\wedge \frac{F}{2\pi}. \ee
 Now let us include the monopole worldline $\ell$ and replace $M$ with the manifold with boundary $M'$.  We consider the same integral as in eqn. (\ref{flimx}) but now on $M'$:
\be\label{dono} \int_{M'} \frac{G}{2\pi}\wedge \frac{F}{2\pi}. \ee
Let us try to define this, as before, by trivializing $w_2(TM')$ along $\partial M'$. The advantage of the formalism we are using here is that we can give a
more direct explanation of what this means.   Trivializing $w_2(TM')$ along $\partial M'$ corresponds, in the language we are using now,
to picking $B$ to be a pure gauge along $\partial M'$.   With this restriction, what freedom is there in picking $B$?   Equation (\ref{flimx}) would allow us to change the flux of $G$ in the interior of $M$
only by an even number of flux quanta.   This corresponds to replacing the integral lift $x$ with a different integral lift of $w_2(TM)$, and it does not affect the integral in eqn. (\ref{dono}) mod 2.
However, we are free to change the flux of $G$ near $\partial M'$ by an odd number of flux quanta.   To be precise, factorize $M'$ near its boundary as $(\ell\times \R_+)\times S^2$,
where $\R_+$ is a half-line (whose endpoint is on $\partial M'$).    Without affecting eqn. (\ref{zimx}) or the fact that $B$ is pure gauge along $\partial M'$, we can change $G$ to $G+G'$
where
\be\label{wono}\int_{\ell\times \R_+}\frac{G'}{2\pi}=1.   \ee
This is the operation that corresponds to changing the trivialization of $w_2(TM')$ near $\partial M'$.   Using eqn. (\ref{wolbo}) again, we see that shifting $G$ to $G+G'$ shifts the integral
(\ref{dono}) by 1.  This is the effect that corresponds to the monopole being a fermion.

 \subsection{Symmetry Breaking To $\Z_2$}\label{gsbtwo}
 
 In section \ref{gsbone}, we used a Higgs triplet to break SU(2) to U(1), thereby reducing an anomalous $(\Spin(4)\times \SU)/\Z_2$ theory to an anomalous $\Spin_c(4)=(\Spin(4)\times \mathrm{U}(1))/\Z_2$
 theory.    We can go further and add a second Higgs triplet, making it possible to break U(1) to $\Z_2$, the center of SU(2).   This gives a theory that at low energies can be viewed locally
 as a $\Z_2$ gauge theory.   But globally, it is much more accurate to say that the effect of the Higgsing from SU(2) to $\Z_2$ is to 
reduce  $(\Spin(4)\times \SU)/\Z_2$ to $(\Spin(4)\times \Z_2)/\Z_2=\Spin(4)$. 
 
 So the low energy theory can have fields in all possible representations of $\Spin(4)$.  In other words, the low energy theory has dynamically generated a spin structure, which is
 determined by the Higgs expectation values.  Though the low
 energy theory can be described locally as a $\Z_2$ gauge theory, globally it is more accurately described as a theory in which one has to sum over the possible choices of a dynamically generated
 spin structure.
 
 Since the underlying anomaly in the original SU(2) description is trivialized once a spin structure is chosen, there is no mystery about the consistency of this low energy phase:
the dynamical generation of a spin structure eliminates the anomaly.

\begin{figure}
 \begin{center}
   \includegraphics[width=2.5in]{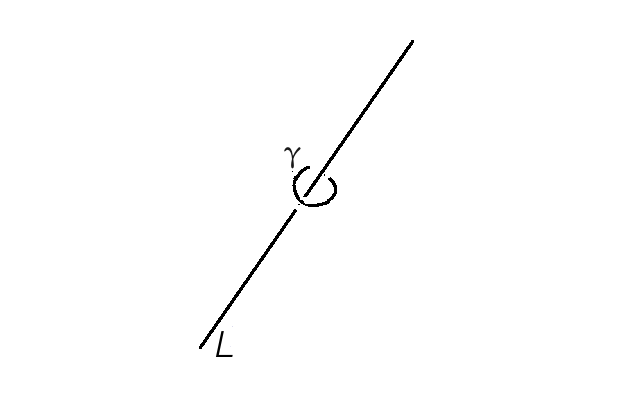}
 \end{center}
\caption{\small   A codimension two vortex line $L$, around which a spin structure does not extend.   A fermion that is parallel transported around a small loop $\gamma$ that links
with $L$ will come back with the opposite sign. The worldvolume of $L$ is a two-manifold $C$, discussed in the text.  \label{Linking}}
\end{figure}

It is interesting to consider the vortex line of the low energy $\Z_2$ gauge theory.   Its worldvolume is a codimension two submanifold $C\subset M$, with the property that the spin structure of $M$
does not extend over $C$; rather, a fermion parallel transported on a tiny loop that is linked with $C$ will change sign (fig. \ref{Linking}).     
In the $\spinc$ description, it is clear how to describe the vortex line: although
the spin structure of $M$ does not extend over $C$ as a spin structure, one can at least locally extend it over $C$ as a $\spinc$ structure.   Physically, this means that in the U(1) description,
there is a half-integral flux in the normal plane to $C$.   An interesting phenomenon occurs if the normal bundle to $C $ in $M$ is unorientable (which is possible even if $M$ itself is orientable).
In that case, there is an obstruction to extending the spin structure of $M$ over $C$
as an everywhere defined
 $\spinc$ structure, because the sign of the U(1)  flux in the normal plane cannot be chosen continuously. There will then have to be a defect line on $C$ along which the sign of the flux changes;
that defect line is actually the worldline of a magnetic monopole.  This monopole is a fermion.
 The $\spinc$ structure cannot be extended over the monopole worldline as a $\spinc$ structure, but it can be extended as a spin-SU(2)
structure.    

\section{Boundary States For A Five-Dimensional Topological Phase}\label{Four}

\subsection{Boundary State Of A 5d Theory With Global SU(2) Symmetry}\label{bg}

As is by now familiar from many other examples, an anomalous fermion theory in four dimensions can be interpreted as the boundary state of a gapped theory in five dimensions.   We will
actually describe two rather different ways to do this.   For the more obvious and standard approach, we consider a five-dimensional gapped theory with fermions and SU(2) global symmetry.
Turn on background SU(2) gauge fields.    There are two cases to consider: a theory that requires a spin structure, and in which the background gauge fields define an SU(2) bundle with connection;
and a theory that can be formulated more generally with a background $\spinSU$ structure.   The old SU(2) anomaly fits in either of these frameworks, but the new anomaly is relevant only
if one considers a theory in which the background fields can define a general $\spinSU$ structure.    (That is because the new anomaly is trivial if a spin structure is assumed, as explained in 
section \ref{nomore}.)

We consider then a theory whose partition function in the presence of these background fields is $(-1)^{\ii_{1/2}}$ or $(-1)^{\ii_{3/2}}$ (or the product of these two).  
Now suppose that $Y$ is a manifold with boundary, and with background fields of the same type (either a spin structure plus an SU(2) bundle with connection, or a $\spinSU$ structure).
Then there is no way to define $\ii_{1/2}$ or $\ii_{3/2}$ as a topological invariant.  To make sense of the spectrum of the Dirac operator $\slashed{D}_5$ on a manifold with boundary, one needs a 
boundary condition.   Here $\slashed{D}_5$ is the five-dimensional Dirac operator coupled to a pseudoreal representation of SU(2); it is an antisymmetric operator, and this is important
in being able to define its mod 2 index.   There is no SU(2)-invariant 
local boundary condition on $\slashed{D}_5$ that preserves  its antisymmetry, and would let us define its
mod 2 index as a topological invariant.  We can use global Atiyah-Patodi-Singer boundary conditions \cite{APS}, which do preserve the antisymmetry,
and with those boundary conditions it is possible to define the mod 2 index.
But the mod 2 index defined with Atiyah-Patodi-Singer boundary conditions is not a topological invariant.   It jumps whenever an eigenvalue  (more precisely, a pair of eigenvalues; see
\cite{WittenRecent}) of the four-dimensional Dirac operator on the boundary
$M=\partial Y$ of $Y$ passes through zero.   This is possible because APS boundary conditions depend explicitly on the spectrum of the Dirac operator on $M$.

This difficulty, however, mirrors the problem in defining a path integral for four-dimensional fermions in a pseudoreal representation of SU(2).   Formally, the fermion path integral is $\Pf(\DD_4)$,
the Pfaffian of the four-dimensional Dirac operator $\DD_4$ (acting on fermions in the appropriate representation of SU(2)).   There is no problem to define the absolute value $|\Pf(\DD_4)|$
of the Pfaffian.    However, there is no simple way to define its sign, and the SU(2) anomalies that we have discussed mean that in some theories, the sign has no satisfactory definition.

The two problems cancel if we combine them.    The absolute value $|\Pf(\DD_4)|$ is not satisfactory for the fermion path integral in the boundary theory, because it gives unphysical behavior
when a pair of eigenvalues of $\DD_4$ pases through zero.     And $(-1)^{\ii(\DD_5)}$ is not satisfactory as the partition function of a gapped theory on a five-manifold with boundary, because
it jumps in sign in an unphysical way when a pair of eigenvalues of $\DD_4$ passes through zero.   However, the product
\be\label{partfn} Z= |\Pf(\DD_4)|(-1)^{\ii(\DD_5)} \ee
is physically sensible and describes a combined system consisting of a gapped but topologically nontrivial system in bulk and a boundary state of gapless fermions transforming in a suitable
representation of SU(2).

This construction of the combined path integral of a bulk system plus a boundary state, 
which is described more fully in \cite{WittenRecent}, has various applications in string/M-theory (see section   2   of \cite{DMW} and section 7.3 of \cite{MW}). 
It has a fairly direct analog (in certain spacetime dimensions) for fermions in a real representation of the gauge group.    For fermions in a complex representation, the construction
has a more subtle analog using the Atiyah-Patodi-Singer $\eta$-invariant \cite{APS} and the Dai-Freed theorem   \cite{DF}, rather than the mod 2 index.    See \cite{WittenRecent} for an introduction
and \cite{WittenDBrane} for an application to the worldsheet path integral of the heterotic string.

\subsection{A Topologically Nontrivial Gapped System Without Symmetry}\label{bsu}

There is a more subtle way to use the new SU(2) anomaly in four dimensions  to give a boundary state for a topologically nontrivial theory in five dimensions.
In this case, we consider a five-dimensional theory with no symmetry assumed.  
Let $J$ be the mod 2 invariant
\be\label{oggo}J(Y)= \int_Y w_2(TY) w_3(TY) \ee
and consider a gapped theory \cite{K,T} whose partition function on a closed five-manifold $Y$  is
\be\label{moggo}(-1)^{J(Y)}. \ee
We want to find a boundary state for this theory.   This is not straightforward because $\int_Y w_2(TY) w_3(TY)$ is not a topological invariant on a manifold with boundary.
To correct for this, there must be some additional structure or some additional degrees of freedom on the boundary.

The quantity $J$ is not just a topological invariant but  a  
{\it cobordism invariant}, meaning that if $Y$ is the boundary of a six-manifold $Z$, then $J=0$.     This is true without any restriction on $Y$ and $Z$ beyond that they are manifolds.
So certainly if $Y$ is assumed to have a $\spinSU$ structure that extends over $Z$, still with $Y=\partial Z$,  then it remains true -- since this is true even without the $\spinSU$ structures
-- that $J=0$.   So we can view $J$ as a cobordism invariant of a five-manifold with $\spinSU$ structure
(a special one that actually does not depend on the $\spinSU$ structure).   But the group of cobordism invariants of a five-manifold with $\spinSU$ structure
is $\Z_2\times \Z_2$ \cite{FH,GPW,WW}, and for generators one can take $\II_{1/2}$ and $\II_{3/2}$, as we discussed in section \ref{nomore}.   So $J$ must be a linear combination of
$\II_{1/2}$ and $\II_{3/2}$.   To determine the coefficients, we just need to evaluate $J$ for the two examples $Y_1=S^4\times S^1$ and $Y_2=\CP^2\rtimes S^1$ that were discussed in section \ref{nomore}.  In fact $J(S^4\times S^1)=0$, since all Stiefel-Whitney classes of $S^4\times S^1$ vanish.
But $J(\CP^2\rtimes S^1)=1$.   

To understand that last statement, let $x$ be the nonzero element of $H^2(\CP^2,\Z_2)=\Z_2$, and $y$ the nonzero element of $H^1(S^1,\Z_2)=\Z_2$.
The cohomology ring of $\CP^2\rtimes S^1$ is generated by $x$ and $y$ with relations $x^3=y^2=0$ and with   $\int_{\CP^2\rtimes S^1}x^2 y=1$.
We have $w_2(\CP^2\rtimes S^1)=x$, basically because $w_2(\CP^2)=x$ and this is unaffected by the presence of the $S^1$.
On the other hand
 $w_3(\CP^2\rtimes S^1)=xy$, leading to $J(\CP^2\rtimes S^1)=1$.    The fact that $w_3(\CP^2\rtimes S^1)=xy$ is related to the following.  As we learned in discussing eqn. (\ref{imx}),
 the obstruction to existence of a $\spinc$ structure on a manifold $Y$ is measured by a class $W_3\in H^3(Y,\Z)$.    $\CP^2\rtimes S^1$ does not admit a $\spinc$ structure,
 as we know from section \ref{spinsu}, so it has $W_3\not=0$.  The Stiefel-Whitney class $w_3$ is defined as the mod 2 reduction of $W_3$.  In the particular case of $Y=\CP^2\rtimes S^1$,
 one has $H^3(Y,\Z)=\Z_2=H^3(Y,\Z_2)$, so nothing is lost in the mod 2 reduction and the fact that $W_3$ is nonzero implies that $w_3$ is also nonzero and is the unique nonzero
 element $xy\in H^3(\CP^2\rtimes S^1,\Z_2)$.   
  
So it follows that  for any five-manifold $Y$ with $\spinSU$ structure, we have
\be\label{noggo}J(Y)=\II_{3/2}(Y). \ee
Since the left hand side of eqn. (\ref{noggo}) does not depend on the $\spinSU$ structure of $Y$, this shows that $\II_{3/2}(Y)$ likewise does not depend on the $\spinSU$ structure.  

A corollary of this result is that the new SU(2) anomaly is never an anomaly in gauge invariance; it is always an anomaly in a combined diffeomorphism plus gauge transformation.   For as remarked
at the end of section \ref{fived}, an anomaly in a gauge transformation on a four-manifold $M$ would correspond to a nonzero mod 2 index on $M\times S^1$, with some $\spinSU$ structure.
But the Stiefel-Whitney classes of $M\times S^1$ are pulled back from $M$, so $\int_{M\times S^1}w_2w_3=0$.  

\subsection{Boundary State With Emergent SU(2) Gauge Symmetry}\label{bemtwo}

The relation (\ref{noggo}) suggests that the bulk theory with partition function $(-1)^J$ could have a boundary state with an emergent SU(2) gauge symmetry coupled to a single
multiplet of fermions of isospin 3/2 (or any other set of fermions that carries the same anomalies).   This is true, but the explanation involves a few ideas beyond what we needed in section
\ref{bg}.

Suppose first that there is some manifold $\t Y$ such that the $\spinSU$ structure of $M$ extends over $\t Y$.   We do not assume that $\t Y$ is isomorphic to $Y$, and it is not necessary to
know if the $\spinSU$ structure of $M$ extends over $Y$.     We orient $\t Y$ so that it can be glued to $Y$ along $M$ to make a smooth manifold $\h Y$.
Then,
denoting the $\spinSU$ structure of $M$ as $A$,  we define the following substitute for the product of $(-1)^{J(Y)}$ and the fermion path integral $\Pf(\DD_M)$:
\be\label{oxo}Z_{M,A,Y}= (-1)^{J(\h Y)} (-1)^{\ii_j(\DD_{\t Y})} |\Pf(\DD_M)|. \ee
Here $\DD_{\t Y}$ is the five-dimensional Dirac operator\footnote{Since there are now different five-manifolds in play, instead of just denoting a five-dimensional Dirac operator as $\DD_5$, we write $\DD_{\t Y}$, etc., indicating the
five-manifold.  To specify the $\spinSU$ structure, we sometimes write $\DD_{\t Y,A}$, etc. } on $\t Y$ for isospin $j$, and $\ii_j(\DD_{\t Y})$ is its mod 2 index with APS boundary conditions (we assume that the isospin $j$ is of the form
$4r+3/2$, so that the fermions under discussion do carry the relevant anomaly).
The point of this definition is two-fold.   First, it is physically sensible, for the same reason as eqn. (\ref{partfn}):
whenever a pair of eigenvalues of $\DD_M$ passes through 0, so that the fermion amplitude should change sign, such a sign change occurs because of a jump in
$\ii_j(\DD_{\t Y})$.    Second, although the factors $(-1)^{J(\h Y)}$ and $(-1)^{\ii_j(\DD_{\t Y})}$ on the right hand side both depend on the choice of $\h Y$, this dependence
cancels out in the product.   Therefore, $Z_{M,A,Y}$ only depends on $M, A,$ and $Y$, that is on $Y$ and the $\spinSU$ structure of $M=\partial Y$, as it should.

\begin{figure}
 \begin{center}
   \includegraphics[width=3.5in]{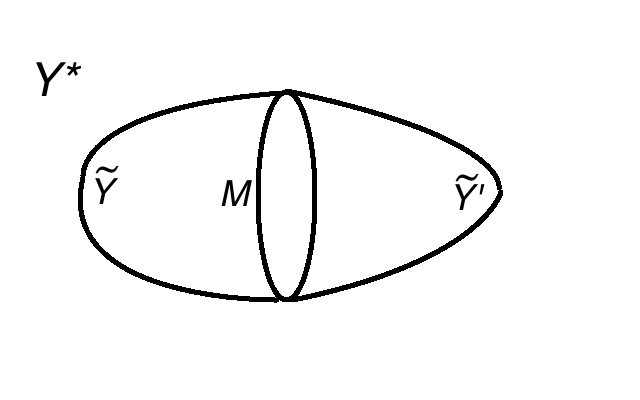}
 \end{center}
\caption{\small   $Y^*$ is obtained by gluing $\t Y$ and $\t Y'$ along their common boundary $M$.   The mod 2 index is additive in such a gluing, leading to eqn. (\ref{woxo}). \label{Multiplicative}}
\end{figure}

\begin{figure}
 \begin{center}
   \includegraphics[width=3.5in]{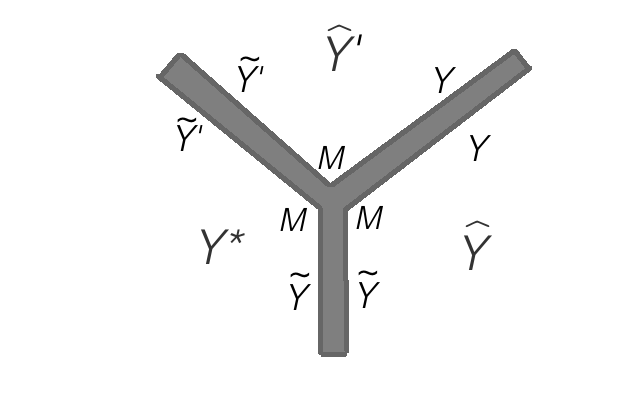}
 \end{center}
\caption{\small   The manifold $\h Y$ is made by gluing together $Y$ and $\t Y$ along their common boundary $M$, and similarly $\h Y'$ is made by gluing together $Y$ and $\t Y'$
along $M$, and $Y^*$ is made by gluing $\t Y$ and $\t Y'$ along $M$.    The shaded region schematically represents a six-manifold $Z$ whose boundary is the union of 
$\h Y$,  $\h Y'$ and $Y^*$.  The existence of this cobordism establishes eqn. (\ref{muyg}).
 \label{Cobordism}}
\end{figure}

To prove the cancellation of the dependence on $\h Y$, we proceed as follows.  Let $\t Y'$ be some other five-manifold of boundary $M$ over which the $\spinSU$ structure of $M$
extends.  Glue $\t Y'$ to $Y$ along $M$ to make a new five-manifold $\h Y'$, and glue $\t Y'$ to $\t Y$ to make another new five-manifold $Y^*$.
  The gluing law for the mod 2 index says that it is additive under gluing, in the sense that, if $Y^*$ is made by gluing $\t Y$ and $\t Y'$ as in fig. \ref{Multiplicative}, 
  then\footnote{The gluing theorem can be proved by picking on $M$ a $\spinSU$ structure such that the Dirac
operator $\DD_M$ has no zero-modes, and then picking on $Y^*$ a metric that near $M$ looks like a long tube $M\times \R$.    In this limit, the low-lying modes of the Dirac
operator on $Y^*$ come from what in the limit of an infinitely long tube would be zero-modes  localized on either $\t Y$ or $\t Y'$, and this leads to the additivity of the mod 2 index.}   
\be\label{woxo} (-1)^{\ii_j(\DD_{\t Y})}(-1)^{\ii_j(\DD_{\t Y'})}=(-1)^{\ii_j(\DD_{Y^*})}.\ee
On the other hand, from eqn. (\ref{noggo}), we have 
\be\label{xitgo}
(-1)^{\ii_j(\DD_{Y^*})}=(-1)^{J(Y^*)}. \ee 
We also have an gluing law for integrals in mod 2 cohomology that is analogous to eqn. (\ref{woxo}):
\be\label{muyg} \int_{\h Y} w_2(T\h Y)w_3(T\h Y)=\int_{\h Y'} w_2(T\h Y')w_3(T\h Y')+ \int_{Y^*}w_2(TY^*)w_3(TY^*), \ee
or simply $J(\h Y)=J(\h Y')+J(Y^*)$.    This statement would follow from elementary calculus if $w_2$ and $w_3$ were differential forms; one would write the integral over $\h Y$
as the sum of an integral over $Y$ and an integral over $\t Y$, and after making a similar decomposition of each term on the right hand side of eqn. (\ref{muyg})
one would find that all terms cancel in pairs.   This
sort of formula is also valid for integration in mod 2 cohomology.    One way to prove it is to observe (fig. \ref{Cobordism}) that there is a fairly obvious six-manifold
$Z$ whose boundary is the union of  $\h Y$, $\h Y'$, and $Y^*$.    The existence of $Z$ implies that $0=\int_{\partial Z} w_2(T\partial Z) w_3(T\partial Z)$, and writing this out in detail
we get eqn. (\ref{muyg}).   (The signs in that formula are not important as these quantities are all $\Z_2$-valued. They have been chosen to make the formulas look natural in ordinary
calculus.)    Using (\ref{woxo}), (\ref{xitgo}),  and (\ref{muyg}), it is straightforward to show that the right hand side of eqn. (\ref{oxo}) does not change if $\t Y$ is replaced by $\t Y'$ and correspondingly
$\h Y$ is replaced by $\h Y'$.   Thus the right hand side of this formula only depends on $Y$ and $A$, as it should.  

There is still a gap in this construction of a boundary state, because we assumed that there exists some $\h Y$ with boundary $M$ over which the $\spinSU$ structure of $M$ extends.
In general, this is not the case.  If $M$ is an abstract four-manifold with $\spinSU$ structures, there are two obstructions to existence of a manifold $\h Y$ with boundary $M$ over
which the $\spinSU$ structure of $M$ extends.   First, the signature of $M$ is an obstruction to existence of $\h Y$, even if one does not ask to extend the $\spinSU$ structure of $M$ over
$\h Y$.  In our case, this obstruction vanishes, since $M$ is given as the boundary of $Y$.   The second obstruction is that if the SU(2) instanton number of $M$ is nonzero, then, regardless of
what $\h Y$ we choose with boundary $M$, the $\spinSU$ structure of $M$ will not extend over $\h Y$.   The SU(2) instanton number normalized to be integer-valued for an ordinary SU(2) gauge
field is
\be\label{tofu}Q=\int_M \frac{\Tr \,F\wedge F}{8\pi^2}.\ee
In gauge theory, one usually introduces an angular parameter $\theta$ and includes a factor $\exp(\i\theta Q)$ in the integrand of the path integral.   As $Q$ is integer-valued, $\theta$ has period $2\pi$.
On a $\spinSU$ manifold $M$, in general $Q$ is not an integer, but if $M$ has zero signature (as is the case if it is the boundary of some $\h Y$), then $Q$ is integer-valued.\footnote{One way
to prove this is to use the index theorem for the index of the $\spinSU$ Dirac operator on $M$ in the spin 1/2 representation of SU(2).   This theorem writes the index, which is an integer,
as the sum of $Q$ and a multiple of the signature of $M$.  But the signature vanishes if $M$ is a boundary, so  $Q$ is an integer, the index.}    
Let us pick a particular example consisting of a five-manifold $Y_0$, a four-manifold $M_0=\partial Y_0$,
and an SU(2) gauge field $A_0$ on $M_0$ with $Q=1$.  For  instance, 
we can pick $M_0$ to be a four-sphere $S^4$ with some metric and some SU(2) gauge field $A_0$ of instanton number 1, and pick $Y_0$ to be a five-ball $B_5$ whose boundary is $M_0$.
We do not have a natural way to define the sign of  $(-1)^{J(Y)} \Pf(\DD_{M_0,A_0})$. We arbitrarily declare this object to be positive and so write 
\be\label{wofu} Z_{M_0,A_0,Y_0}= e^{\i\theta}|\Pf(\DD_{M_0,A_0})|. \ee
We can regard this as the definition of what we mean by $\theta$ at the quantum level.  If we took
$(-1)^{J(Y)} \Pf(\DD_{M_0,A_0})$  to be negative, this would have the same effect in eqn. (\ref{wofu}) and later as shifting $\theta$ by $\pi$.

\begin{figure}
 \begin{center}
   \includegraphics[width=3.5in]{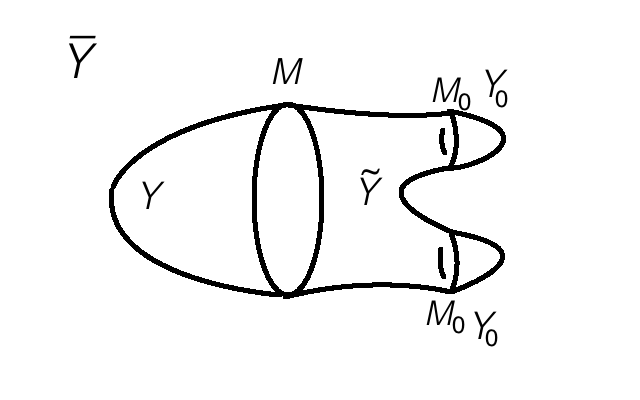}
 \end{center}
\caption{\small   $M$ is cobordant to $n$ copies of $M_0$, for some $n$, via a five-manifold $\t Y$ over which the $\spinSU$ structures of $M$ and $M_0$ extend.   This is depicted
here for $n=2$.   Each copy of $M_0$ is the boundary of some $Y_0$, and $M$ is the boundary of $Y$.   So $Y$, $\t Y$, and $n$ copies of $Y_0$ glue together naturally to a closed five-manifold
$\bar Y$, depicted here.   This construction is used in the general definition of the amplitude $Z_{M,A,Y}$.
 \label{ComplexCobordism}}
\end{figure}

Now we want to define $Z_{M,A,Y}$ for an arbitrary triple $M,A,Y$.   If the instanton number on $M$ is a nonzero integer $n$, then we cannot directly use eqn. (\ref{oxo}) to define
$Z_{M,A,Y}$, because there is no $\t Y$ with boundary $M$ over which the $\spinSU$ structure of $M$ extends.   But there is a $\t Y$ with boundary\footnote{Here $M-nM_0$ is the disjoint union
of $M$ with $|n|$ copies of $M_0$; if $n>0$, the minus sign in $M-nM_0$ means that one reverses the orientation of $M_0$.} $M-nM_0$  over which the $\spinSU$ structures of $M$ and $M_0$ extend.  
Finally, by gluing together $Y$, $\t Y$, and $n$ copies of $Y_0$ along their boundaries -- which together consist of $M$ and $n$ copies of $M_0$ -- we build a closed five-manifold $\overline Y$,
depicted in fig. \ref{ComplexCobordism}.   The general definition of  $Z_{M,A,Y}$ is then
\be\label{nofu} Z_{M,A,Y}   = e^{\i n\theta} (-1)^{J(\bar Y)} (-1)^{\ii_j(\DD_{\t Y})} |\Pf(\DD_{M,A})|. \ee
As usual, this definition is physically sensible in part because whenever an eigenvalue pair of $\DD_{M,A}$ passes through zero, the desired jump in sign of the amplitude
comes from the behavior of the mod 2 index $\ii_j(\DD_{\t Y})$.   Beyond this, the consistency of eqn. (\ref{nofu}) amounts to two facts.  First, the right hand side does not depend on the choice of
$\t Y$.   This follows from the same reasoning that we used in discussing eqn. (\ref{oxo}).   Second, if we take $M=M_0$, $A=A_0$, $Y=Y_0$, then eqn. (\ref{nofu}) is compatible with eqn. (\ref{wofu}).
To show this, we observe that if $M=M_0$, then $n=1$ and 
we can take $\t Y$ to be $M_0\times I$ (where $I$ is a unit interval) with a $\spinSU$ structure pulled back from $M_0$.  Then $J(\bar Y)=\ii_j(\DD_{\t Y})=0$
and the two formulas for $Z_{M_0,A_0,Y_0}$ coincide.

At this level of description, we have found a boundary state with an emergent SU(2) gauge field coupled to masssless fermions.   Next we should discuss the dynamics of this SU(2) gauge theory.
If the fermion representation is large enough so that the SU(2) gauge theory is infrared-free, then the SU(2) dynamics will not qualitatively change the picture.
For example, if  the anomaly
is carried by a boundary fermion of isospin $4r+3/2$, $r>0$, then the one-loop beta function of the SU(2) gauge theory is positive and, if the SU(2) coupling is small enough, it will flow to zero in the
infrared.   However, for a minimal set of boundary fermions -- a single multiplet of isospin 3/2 --
the one-loop beta function of the SU(2) gauge theory is slightly negative.   The SU(2) dynamics 
cannot be ignored in this case, and the outcome is not clear.   It is difficult to decide, for example, if the theory flows to an infrared fixed point, or to a confining state with a mass gap.  
An analysis leading to confinement was made in \cite{UP}.
A possibility,
suggested by the most attractive channel (MAC) hypothesis \cite{RDS}, is that the theory might Higgs itself to U(1) or possibly to $\Z_2$.   The logic here is that for this particular theory,
the most attractive channel for fermion condensation (assuming Lorentz invariance is maintained) is the isospin 1 channel.   A fermion condensate in this channel will break the gauge symmetry to
U(1) or $\Z_2$, depending on whether the condensate can be made real by a chiral U(1) rotation of the fermions.

\subsection{Boundary State With U(1) Gauge Symmetry}\label{bu}

Once we have found a boundary state with emergent SU(2) gauge symmetry, it is clear that it is possible to get a boundary state with emergent U(1) gauge theory.
We simply follow the same symmetry-breaking as in section \ref{gsbone}: we add a Higgs field $\vec\phi$ of isospin 1, whose expectation value breaks SU(2) to U(1).

After symmetry breaking, the isospin 3/2 fermions that contribute the anomaly in the SU(2) description can gain mass, so the anomaly
must be carried by effective couplings of the U(1) gauge field.
How this happens was already analyzed in sections \ref{gsbone} and \ref{allfer}: we get the anomalous ``all-fermion electrodynamics'' in which particles carrying a unit of electric or magnetic
charge are fermions.    In this theory, since there are no massless charged fermions, the anomaly is carried by effective couplings of the U(1) gauge fields only.   From our previous
analysis, we already know how this happens.   To make the charge 1 monopole a fermion, the integrand of the path integral has   a factor \be\label{impor} (-1)^{\int_M w_2(TM)\cdot c_1(\L)},\ee
where $\L$ is the line bundle on which the U(1) gauge field is a connection.   To make a particle of electric charge 1 a fermion, the ``U(1) gauge field'' is really a $\spinc$ connection.
Together these two conditions are incompatible, leading to an anomaly.  

Thus, it follows by Higgsing from our construction in section \ref{bsu} that an emergent U(1) theory of all-fermion electrodynamics gives a possible boundary state\footnote{This boundary state is already known \cite{T,McGreevy}.  The underlying anomaly, viewed as a mixed anomaly between 1-form global symmetries in
four-dimensional U(1) gauge theory, was described in \cite{OneForm}, and the use of this anomaly to construct this boundary state was briefly indicated in footnote 31
of \cite{CD}.}
 for the five-dimensional
theory with partition function $(-1)^J$.   
The boundary state with emergent all-fermion electrodynamics can be described in the following way, more in the spirit of  \cite{WWW}.   The problem we are facing, which is to define what
we mean by $J(Y)=\int_Y w_2(TY)w_3(TY)$ when $Y$ has a boundary $M$, is analogous to the problem that we already discussed in section \ref{detail}, where we were trying to define
the integral of a different cohomology class (namely $w_2(TM') \cdot c_1(\L)$) on a manifold $M'$ with boundary.   As in that case, a trivialization of the cohomology class along the boundary
gives a way to define the integral $J(Y)$.

We can trivialize $w_2(TY) w_3(TY)$ along the boundary of $Y$ by trivializing either $w_2(TY)$ or $w_3(TY)$ along the boundary.   For our present application, we want to trivialize $w_3(TY)$
along the boundary.  Physically, if we assume the existence of an emergent $\spinc$ connection on $M=\partial Y$, this gives a natural mechanism to trivialize $W_3(Y)$ and hence
$w_3(Y)$ along $M$.    With this assumption, let us see how to describe the boundary state.      Suppose first that there is some manifold $\t Y$ with boundary $M$ such that the $\spinc$
connection $A$ of $M$ extends over $\t Y$.   We do {\it not} assume that $\t Y$ is isomorphic to $Y$, and in general $A$ does not extend over $Y$.   The allowed choices of $\t Y$
depend on $A$, because of the condition that $A$ must extend over $\t Y$.
We orient $\t Y$ so that $\t Y$ and $Y$ can be glued together along $M$ to make a closed five-manifold $\h Y$.   Then we define
\be\label{uyg} J_A(Y) = J(\h Y). \ee
The right hand side is not necessarily zero, because $\h Y$ may not admit a $\spinc$ connection (only $\t Y$ is guaranteed from the construction to admit one), and it may depend
on the original $\spinc$ connection $A$ on $M$ because this affects the allowed choices of $\t Y$.    The point of the definition in eqn. (\ref{uyg}) is, however, that the right hand side
does not depend on the choice of $\t Y$, only on the original five-manifold $Y$ with the emergent $\spinc$ connection $A$ on its boundary.   To see this, let $\t Y'$ be some other five-manifold
with boundary $M$ over which $A$ extends, and let $\h Y'$ be made by gluing together $Y$ and $\t Y'$.   Also, let $Y^*$ be made by gluing together $\t Y$ and $\t Y'$ along $M$, after
reversing the orientation of one so that they fit smoothly.   Then we have the identity (\ref{muyg}) that we used previously in a similar context.   
But now we have the further simplification that 
$Y^*$ is a $\spinc$ manifold, since it was defined by gluing together $\t Y$ and $\t Y'$, over each of which the $\spinc$ structure of $M$ extends.    So $w_3(TY^*)=0$
and therefore the second term on 
 the right hand side of eqn. (\ref{muyg}) vanishes.    The identity reduces to $ \int_{\h Y} w_2(T\h Y)w_3(T\h Y)= \int_{\h Y'} w_2(T\h Y')w_3(T\h Y')$, showing  that $J_A(Y)$ as
defined in eqn. (\ref{uyg}) does not depend on the choice of $\t Y$.

This enables us to define the boundary state.   As a substitute for $(-1)^{J(Y)}$, which is the partition function in the absence of a boundary, we include in the path integral over $A$
a factor $(-1)^{J_A(Y)}$.   This makes sense because as we have seen this factor only depends on $A$ and $Y$.   In addition to this factor, the integrand of the path integral contains
some factor $\exp(-I(A))$ where $I(A)$ is a conventional local action; for example we certainly expect a Maxwell term $\frac{1}{4e^2}\int_M \d^4x \sqrt g F_{ij}F^{ij}$, where $F=\d A$.  
(We also expect a topological term $\theta\int_M F\wedge F/(2\pi)^2$, about which more later.)
An important detail, however, is that we do {\it not} add the factor $(-1)^{\int_M w_2(TM) \cdot c_1(\L)}$ that was important in section \ref{allfer}.   Rather, this term is automatically
incorporated in the factor $(-1)^{J_A(Y)}$ that we have already included.    How this happens is similar to what we explained in section \ref{detail}.  The definition of $J_A(Y)$ used a $\spinc$
connection $A$ to trivialize $w_3(Y)$ along $M=\partial Y$.   If $A$ and $A'$ are two $\spinc$ connections, then their {\it difference} $A-A'$ is an ordinary U(1) gauge field, a connection on a complex
line bundle $\L^*\to M$.   Reasoning similar to that in section \ref{allfer} leads to 
\be\label{kuyp}(-1)^{J_{A'}(Y)}=(-1)^{J_A(Y)} (-1)^{\int_M w_2(TM)\cdot c_1(\L^*)}. \ee
Informally, the last factor on the
right hand side is $(-1)^{\int_M w_2(TM)\cdot (F/2\pi - F'/2\pi)}.$    Keeping $A'$ fixed and letting $A$ vary, we get, up to a constant multiple (that is, an additive constant
in the exponent), the 
interaction $(-1)^{\int_M w_2(TM) \cdot F/2\pi}$ that was discussed in section \ref{allfer}.   The reason that we recover the previous result only up to a constant multiple is that in the present
discussion, we are studying not just all-fermion electrodynamics on the boundary but its combination with a bulk topological field theory.

\def\top{{\mathrm{top}}}
In the above discussion, we assumed that $M$ is the boundary of some manifold $\t Y$ over which the $\spinc$ structure of $M$ extends.   In general, this is not the case.  
Just as in section \ref{bsu}, this means not that the boundary state does not exist but that it is subtle to describe in this state what we mean by the topological angle $\theta$.  
The obstruction to existence of $\t Y$ (given that $M$, ignoring its $\spinc$ structure, is the boundary of $Y$) is that the $\spinc$ connection on $M$ might have a nonzero
instanton number.  The integer-valued instanton number in U(1) gauge theory is 
\be\label{worbo}Q=\int_M \frac{F\wedge F}{(2\pi)^2}. \ee
In gauge theory, one usually includes in the path integral a factor $\exp(\i\theta Q)$.  As $Q$ is an integer,
it follows that 
in purely four-dimensional gauge theory, $\theta$ is an angular variable with period $2\pi$.  
In the present context, matters are somewhat different.   Given that $M$ is the boundary of some $Y$, $Q$ takes values in\footnote{We will
explain the proof somewhat informally.   The flux condition (\ref{welgo}) means that we can think of $F/\pi$ as $w_2+2x$, where $x$ is an integral class, so $Q=\frac{1}{4}\int_M(w_2^2+4w_2x+4x^2)$.
Since $\int_M w_2^2$ is a cobordism invariant and $M=\partial Y$ is a boundary, this term vanishes and $Q=\int_M(w_2x+x^2)$.    A standard result is that for any integral class $x$,
$\int_M w_2x$ is congruent mod 2 to $\int_M x^2$, and hence $Q$ is congruent mod 2 to $2\int_M x^2$ and is an even integer.}
$2\Z$, so naively $\theta$ has period $\pi$.   But there is a subtlety in defining what we mean by $\theta$, related to the fact that 
 if $Q\not=0$, then $\t Y$ does not exist.   We proceed as in section \ref{bemtwo}.    Pick an example with $Q=2$, the smallest possible value; for instance, 
 one can take $M_0=S^2\times S^2$, with a $\spinc$ connection $A_0$, of field strength $F_0$,
such that $F_0/2\pi$ integrates to 1 over either $S^2$ factor in $M_0$.   $M_0$ is the boundary of, for example, $Y_0=B_3 \times S^2$, where $B_3$ is a three-ball with boundary $S^2$.  
(The $\spinc$ structure of $M_0$ does not extend over $Y_0$, since this would contradict the fact that $\int_{\partial B_3}F_0/2\pi=\int_{S^2}F_0/2\pi=1$.)   Because we do not have
a natural way to define $(-1)^{J(Y_0)}$ in this example, we arbitrarily declare it to be positive.   This means that the phase of the amplitude for $(Y_0,M_0,A_0)$ is $e^{2\i\theta}$, fixing
what we mean by $\theta$ at the quantum level.    
Once this is specified, we proceed for any example $(Y,M,A)$ in the same way as in section \ref{bemtwo}.
We simply observe that any $(M,A)$ is cobordant to a certain number of copies of $(M_0,A_0)$, and then we proceed as before.   The upshot is that the arbitrarily chosen overall
sign of the path integral amplitude for  $(Y_0,M_0,A_0)$ enables us to define the path integral amplitude for any $(Y,M,A)$.   If the instanton number integrated on $M$ is $n$,
then the path integral amplitude is
\be\label{toldo} Z_{M,A,Y}= (-1)^{J(\bar Y)} e^{\i n\theta} \exp(-I(A,M)),\ee
where $\bar Y$ is defined as in fig. \ref{ComplexCobordism}.  
In eqn. (\ref{toldo}) $I(A,M)$ is the action for the $\spinc$ connection on $M$, except that we omit the $\theta$ term in $I$ and write it explicitly as a factor $e^{\i n\theta}$.   The reason for doing this is that 
it is subtle to explain what we mean by $\theta$.
Making a different choice of the reference point $(Y_0,M_0,A_0)$ would change what we mean by $\bar Y$ and thus would possibly shift $J(\bar Y)$.  If so,
this would  
have the same effect as shifting $\theta$ by $\pi/2$.  Likewise, changing the sign of the amplitude at $(Y_0,M_0,A_0)$ would have the effect of shifting $\theta$ by $\pi/2$.
So the parameter $\theta$ that appears in eqn. (\ref{toldo}) is really not just a parameter from the classical action, but is a sort of quantum-corrected  parameter.

\subsection{Boundary State With A Dynamical Spin Structure}\label{bdu}

As in section \ref{gsbtwo}, we can go farther and, by giving an expectation value to a Higgs field of charge 2, we can spontaneously break the U(1) gauge symmetry to $\Z_2$.
To describe the situation more geometrically, this reduces $(\Spin(4)\times \mathrm{U}(1))/\Z_2$ to $(\Spin(4)\times \Z_2)/\Z_2=\Spin(4)$ and means that the emergent
$\spinc$ connection is reduced to an emergent spin structure.  

There is a fairly evident reason that such a boundary state is possible.   An emergent spin structure on $M=\partial Y$ trivializes $w_2(TY)$ near the boundary and therefore,
topologically, makes it possible to define what we mean by $J(Y)=\int_Y w_2(TY) w_3(TY)$. 

\subsection{Another Boundary State?}\label{onemore}

At least one more boundary state of this theory can be described in a somewhat similar way.   By Higgsing from SU(2) to U(1), or by directly postulating the existence on $M=\partial Y$
of an emergent $\spinc$ connection, we described a boundary state in which $W_3(Y)$ is trivialized along the boundary.   This implies a trivialization of $w_3(Y)$,
and made it possible to define $J(Y)$.

Existence of a $\spinc$ connection does not in general lead to an arbitrary trivialization of $w_3(Y)$, and of course a $\spinc$ connection (since it looks locally like a U(1) gauge field)
describes much more than a trivialization of $w_3(Y)$.   A boundary state defined with an emergent $\spinc$ connection is gapless.

We could instead have a gapped boundary state in which the emergent variable along the boundary $M$ is a trivialization of $w_3(M)$ -- nothing more nor less.  Two trivializations
of $w_3(M)$ differ by an element of $H^2(M,\Z_2)$, so such a boundary state looks locally like we are summing over an element of $H^2(M,\Z_2)$, though the global structure
is more subtle.

It is not completely clear if this boundary state is  new or is a dual formulation of the one described in section \ref{bdu}.

\subsection{Trivialization Of the  New SU(2) Anomaly By A Spin Or Spin$_c$ Structure}\label{notenew}

We can also use these ideas to explain in a precise way the statement that  a theory with the new SU(2) anomaly can be formulated in a consistent fashion on a four-manifold
$M$ that is endowed with a choice of a spin or $\spinc$ structure.   Spin and $\spinc$ structures are relevant, because they trivialize $w_2$ and $w_3$, respectively, and therefore trivialize
the anomaly $\int_Y w_2w_3$.

As always, the problem is to explain what is meant by the Pfaffian $\Pf(\slashed{D}_M)$ of the four-dimensional fermion operator $\slashed{D}_M$.    
In the abstract, on a four-manifold $M$ with a $\spinSU$ structure and no further structure,
there is no way to determine the sign of this Pfaffian.  However, suppose that $M$ is endowed with a spin structure or a $\spinc$ structure, and suppose that $M$ is the boundary of a five-manifold
$Y$ over which the spin or $\spinc$ structure of $M$ extends.    Then as in eqn. (\ref{partfn}), we define the fermion path integral on $M$ as
\be\label{toffo}   |\Pf(\slashed{D}_M)|(-1)^{\II(\slashed{D}_Y)}. \ee

By familiar arguments, using the fact that $\int_Y w_2w_3=0$ on a spin or $\spinc$ manifold, we deduce that in the case of a theory that has the new SU(2) anomaly but not the old one,
this definition of the fermion path integral does not depend on the choice of $Y$.    If a suitable $Y$ does not exist, we deal with this as in section \ref{bemtwo} by assigning arbitrary phases
to generators of the appropriate cobordism group.  The group that classifies four-manifolds $M$ with $\spinSU$ structure up to cobordism is $\Z\times \Z$, with the two invariants
being the SU(2) instanton number and the signature of $M$ (which one could call the gravitational instanton number).   After assigning arbitrary phases to generators of the cobordism
group, one gets a definition of the fermion path integral on $M$ along with a definition of the gauge theory and gravitational theta-angles at the quantum level.   But all this depends
on having picked on $M$ a spin or $\spinc$ structure to trivialize the anomaly.

\subsection{Analogs In Four Dimensions}\label{anafour}

Much of what we have said has interesting  analogs for certain topological phases of matter in four dimensions.   The following discussion will be  in the spirit of \cite{WWW}  (and part of the following
was explained there).

In four dimensions, there is a gapped theory of bosons, not protected by any symmetry, whose partition function on a four-manifold $M$ is
$(-1)^{K(M)}$ where $K(M)=\int_M w_2(TM)\cdot w_2(TM)$.    From what we have said, it is clear how to construct a gapped boundary state for this theory.
If $M$ has a non-empty boundary $W$, then $K(M)$ is not well-defined {\it a priori}, but it becomes well-defined if one is given a trivialization of  $w_2(TM)$ near $W$.
Such a trivialization corresponds naturally to a spin structure on $W$.  So this gives a gapped boundary state that locally can be described
in terms of an emergent $\Z_2$ gauge symmetry on $W$, but is better described by saying that the theory has a dynamically generated spin structure.
A spin structure makes it possible to describe the propagation of fermions, so such a theory will generically have fermionic quasiparticles on the boundary, though in bulk it has bosons only.

The theory with partition function $(-1)^{K(M)}$ makes sense on unorientable manifolds, since $K(M)$ is well-defined on a possibly unorientable closed four-manifold.   
Placing this theory on an unorientable manifold is only natural if one assumes time-reversal or reflection symmetry and for brevity we will refer to time-reversal
symmetry.\footnote{In a relativistic theory, the $\sf{CRT}$ theorem relates time-reversal symmetry to
reflection symmetry so either of these implies the other and there is no distinction.  In condensed matter physics, the two cases are potentially different. They lead to the same
physics at low energies when there is an emergent relativistic description (as we assume here).  But in condensed matter physics, $\sf R$ and $\sf T$
can allow different relevant parameters that might potentially obstruct the emergence of a relativistic description at low energies, so the problem of finding an emergent relativistic
description can be essentially different depending on whether a system has microscopic $\sf R$ symmetry or  microscopic $\sf T$ symmetry.}
There are two possible generalizations of a spin structure
to an unorientable manifold, namely a  $\pin^+$ structure and a $\pin^-$ structure, where in the former case, a fermionic quasiparticle will transform as a Kramers
doublet under time-reversal, and in the latter case,  it will transform 
as a Kramers singlet.   Since on an unorientable manifold, $w_2(TM)$ is the obstruction to a $\pin^+$ structure (and so is trivialized by a choice
of $\pin^+$ structure), the dynamical spin structure of the last paragraph becomes a dynamical $\pin^+$ structure once we allow spacetime to be unorientable.
So  once we assume time-reversal symmetry, the theory with partition function $(-1)^{K(M)}$
has a gapped boundary state with dynamical $\pin^+$ structure and with  Kramers doublet fermions on the
boundary.

Once we assume time-reversal symmetry, another invariant is possible, namely $L(M)=\int_M w_1(TM)^4$.   There are consequently two more theories,
namely the ones whose partition functions on a closed manifold are $(-1)^{L(M)}$ and $(-1)^{L(M)+K(M)}=(-1)^{T(M)}$, where $T(M)=\int_M (w_2(TM)+w_1(TM)^2)^2$.
Both of these theories admit gapped boundary states that locally can be described by an emergent $\Z_2$ gauge theory, but globally have a more illuminating description.

We consider first the theory whose bulk partition function is $(-1)^{T(M)}$.  Here, if $M$ has boundary $W$, we need to trivialize $(w_2(TM)+w_1(TM)^2)^2$ near $W$.
An obvious way to do this is to trivialize $w_2(TM)+w_1(TM)^2$ near $W$.   As $w_2(TM)+w_1(TM)^2$ is the obstruction to a $\pin^-$ structure, it is trivialized by the choice of a
$\pin^-$ structure.   Thus this theory has a gapped boundary state with a dynamical $\pin^-$ structure on the boundary.    This boundary state will have Kramers singlet fermionic
quasiparticles.

Finally, we consider the theory with bulk partition function $(-1)^{L(M)}$.   To define $\int_M w_1(TM)^4$ on a four-manifold $M$ with boundary $W$, we need to trivialize $w_1(TM)^4$
near the boundary.   One way to do this is to trivialize $w_1(TM)^2$.  Let us discuss what sort of boundary state this will give.   We follow a logic similar to what we discussed in section
\ref{detail}.   We start with the short exact sequence of abelian groups:
\be\label{nunix} 0\to \Z_2 \overset{2}{\rightarrow} \Z_4\overset{r}{\rightarrow}\Z_2\to 0,\ee
 where the first map is multiplication by 2 and the second is reduction mod 2.
 This leads to a long exact sequence of cohomology groups that reads in part
 \be\label{womx}\cdots \to H^1(M,\Z_4)\overset{r}{\rightarrow} H^1(M,\Z_2) \overset{\beta}{\rightarrow} H^2(M,\Z_2)\to \cdots . \ee
 It can be shown that for any class $x\in H^1(M,\Z_2)$, $\beta(x)=x^2$.  Thus the exactness of the sequence tells us
 that $x^2=0$ if and only if $x=r(y)$ for some $y$, and more to the point that the choice of such a $y$ gives a trivialization of $x^2$.
 
 In our application, we want to think of the sequence as follows:
\be\label{nunox} 0\to \Z_2^{\mathrm{emergent}} \overset{2}{\rightarrow} \Z_4^{\mathrm{total}}\overset{r}{\rightarrow}\Z_2^{\sf T}\to 0.\ee
Here we take $\Z_2^{\sf T}$ to be the $\Z_2$ symmetry generated by time-reversal $\sf T$.  $\Z_2^{\mathrm{emergent}}$ will be  an emergent $\Z_2$ gauge symmetry that
exists only along the boundary.  The combined symmetry group will be  $\Z_4^{\mathrm{total}}$.   We choose $x=w_1(T)$.    $x$ is the first Stiefel-Whitney class of a $\Z_2$ bundle
over $M$ that is known as the orientation bundle of $M$:
 it is a $\Z_2$ bundle $\varepsilon$
whose holonomy around a loop $\ell$ is $+1$ or $-1$ depending on whether the orientation of $M$ is preserved in going around $\ell$.   We have found that we can trivialize $x^2$
by picking a ``lift'' of $x$ to a mod 4 class $y$ whose mod 2 reduction is $x$.   A choice of $y$ corresponds to a $\Z_4$ bundle whose holonomy around a loop $\ell$ is the square
root of the holonomy of $\varepsilon$ (so for an orientation-preserving loop, the holonomy is $\pm 1$, and for an orientation-reversing one, it is $\pm \i$).   In a boundary state
based on emergent trivializations of $w_1(TM)^2$, we sum over such structures.   Two such structures differ by an element of $H^1(M,\Z_2)$, that is, a flat $\Z_2$-valued gauge field,
so the boundary state locally looks like a theory with an emergent $\Z_2$ gauge symmetry, though that is not the best description globally.

Let us consider the quasiparticles in such a boundary state.   The boundary state gives no choice of spin or  $\pin^\pm$ structure, so the quasiparticles are bosons.   Consider
a quasiparticle state $|\Psi\rangle$ that is odd under $\Z_2^{\mathrm{emergent}}$.   The embedding in eqn. (\ref{nunx}) tells us that the nonzero element of this group corresponds to the element 
$2\in \Z_4^{\mathrm{total}}$.   So a quasiparticle state $|\Psi\rangle$ that carries the emergent gauge charge is odd under the element 2 of $ \Z_4^{\mathrm{total}}$, that is,
this element acts  by $|\Psi\rangle\to -|\Psi\rangle$.   We can lift the time-reversal symmetry
$\sf T$ to either generator 1 or 3 of $ \Z_4^{\mathrm{total}}$.  Whichever one we pick, $\sf T^2$ corresponds to the element 2 of $ \Z_4^{\mathrm{total}}$ and, as we have just
seen, acts by $|\Psi\rangle\to -|\Psi\rangle$.   Since $\sf T^2|\Psi\rangle=-|\Psi\rangle$,  the quasiparticles that carry the emergent gauge charge are Kramers doublet bosons.

A $\Z_2$ theory in $2+1$ dimensions has vortex particles as well as particles carrying the $\Z_2$ charge.  By extending the reasoning, it is possible to show that in each of the three
examples, the vortex has the same properties (boson or fermion, Kramers singlet or doublet) as the charge.

For the theory with partition function $(-1)^{L(M)}$, we can also construct a boundary state in which the emergent variable on the boundary $W$ is a trivialization of $w_1(TW)^3$.
This can be discussed analogously to what we said in section \ref{onemore}.   One might wonder what is a boundary state with an emergent variable that trivializes $w_1(TW)$, rather
than one of its powers.   This corresponds to a state in which $\sf T$ is spontaneously broken, since a field whose expectation value
  trivializes $w_1(TW)$, while preserving Poincar\'e symmetry,
is a $\sf T$-odd scalar field.

\section{Some Generalizations}\label{Five}

What we have described in this paper has some generalizations that are worth  at least brief mention.  

\subsection{Different Gauge Groups}\label{diffgroups}

\subsubsection{Sp(2N)}

First of all, one can replace SU(2) with another gauge group.   The closest analog is\footnote{Our terminology is such that Sp(2) coincides with SU(2), and in general the fundamental
representation of Sp(2N) has complex dimension 2N.}  Sp(2N), which of course has pseudoreal representations just like SU(2).
Following the logic of section \ref{Two}, it is  not difficult to see that the original SU(2) anomaly has a close analog for a single multiplet of four-dimensional fermions
in the fundamental representation of Sp(2N).   Moreover, the new SU(2) anomaly has a close analog for a single multiplet in the third symmetric tensor product of the fundamental
representation of Sp(2N).    Both of these statements can be generalized to arbitrary Sp(2N) representations by repeating what we have done for SU(2), but we will not
explore that direction.

\subsubsection{SO(3)$\times$SU(2)}\label{SOS}

An SO(3) gauge theory with SU(2) global symmetry and a single multiplet of fermions transforming as $(1,1/2)$ under $\SOS$ (that is, isospin 1 under SO(3) and isospin 1/2 under
SU(2)) was recently studied in \cite{CD}.   It was shown that interesting anomalies arise if one gauges the SU(2) global symmetry and then formulates the theory
on a $\spinSU$ manifold.     There are three possible cobordism invariants, which
one can take to be $\II_{1/2}$, $\int_Y w_2(TY)w_3(TY)$, and $\int_Y w_2(E) w_3(TY)$, where $E$ is the SO(3) bundle.    It was found that the anomaly of this theory
is $\II_{1/2}+\int_Y w_2(E) w_3(TY)$.   The term $\int_Y w_2(E) w_3(TY)$ was found by reinterpreting a computation \cite{WittenS} in a topologically twisted supersymmetric gauge theory.

From the point of view of the present paper, one would say that since the $(1,1/2)$ representation of $\SOS$ is pseudoreal, the anomaly for this representation is given by the mod 2 index $\II_{1,1/2}$
of the Dirac operator with values in this representation.    To show that $\II_{1,1/2}=\II_{1/2}+\int_Y w_2(E)w_3(TY)$, it suffices to calculate for a set of manifolds that detect the three invariants.
For such a set, we can take $Y_1=S^4\times S^1$, with an SU(2) bundle of instanton number 1 on $S^4$ and a trivial SO(3) bundle; $Y_2'=\CP^2\rtimes S^1$, with a $\spinSU$ structure of flux 1/2 and trivial SO(3) bundle;
and $Y_2''=\CP^2\rtimes S^1$, with a $\spinSU$ structure of flux 1/2 and an SO(3) bundle whose restriction to $\CP^2$ has $w_2(E)\not=0$.    One can compute $\II_{1,1/2}$ for these three
examples,
using facts described in section \ref{Two} ,
 and (using also the determination of $w_3(TY)$ in section \ref{bsu})
 confirm that $\II_{1,1/2}=\II_{1/2}+\int_Y w_2(E)w_3(TY)$.    The left and right hand sides are both nonzero for $Y_1$ and $Y_2''$ and zero for $Y_2'$.

After suitable Higgsing to U(1), an SO(3)$\times$SU(2) gauge theory with fermions in the $(1,1/2)\oplus (0,1/2)$ representation gives an ultraviolet completion of all-fermion electrodynamics.
(The role of the $(0,1/2)$ fermions is to cancel the old SU(2) anomaly.)   For this, one first  breaks the symmetry from SO(3)$\times$SU(2) to a diagonally embedded SU(2) subgroup.
Under this subgroup, the fermions transform as the direct sum of isospin 3/2 and two copies of isospin 1/2, so further symmetry breaking to U(1) will give a model of all-fermion
electrodynamics.

\subsubsection{Grand Unification Without Spin Structure}

An interesting case that does {\it not} lead to a new anomaly is the following.   One of the standard Grand Unified Theories of four-dimensional particle physics is a $\Spin(10)$ gauge theory in which
fermions are in spinor representation of $\Spin(10)$ and bosons are in tensor representations.   Thus fermions are odd under a $2\pi$ rotation in $\Spin(10)$, while bosons are even.
Hence this theory can potentially be formulated on a manifold with $(\Spin(4)\times \Spin(10))/\Z_2$ structure.  

This theory is completely anomaly-free.   By standard criteria, the model
lacks perturbative anomalies.   One way to show that it has no global anomaly is to first observe that the only cobordism invariant of a five-manifold $Y$ with $(\Spin(5)\times \Spin(10))/\Z_2$
structure is $\int_Y w_2(TY)w_3(TY)$ (which does not really depend on the $(\Spin(5)\times \Spin(10))/\Z_2$ structure).   To check that the $\Spin(10)$ Grand Unified Theory has no anomaly
associated to the invariant  $\int_Y w_2(TY)w_3(TY)$, it suffices to consider an example.  In section \ref{Two}, we constructed a $(\Spin(4)\times \SU)/\Z_2$ structure on $\CP^2$
and found an anomaly because the path integral measure was not invariant under a certain diffeomorphism plus gauge transformation $\h\varphi$.  Since $\SU$ is isomorphic to $\Spin(3)$,
which embeds in $\Spin(10)$ in an obvious way, we can promote this example to a $(\Spin(4)\times \Spin(10))/\Z_2$ structure on $\CP^2$.    Now recall that in this example,
the $(\Spin(4)\times \SU)/\Z_2$ Dirac operator acting in the isospin 1/2 representation of $\SU$ has no zero-modes.    The spinor representation $\mathbf{16}$ of $\Spin(10)$ (the usual
representation for Grand Unification) decomposes under $\SU\cong \Spin(3)$ as the direct sum of eight copies of the isospin 1/2 representation.  So it too has no zero-modes and hence
the path integral measure is invariant under any classical symmetry of the bosonic background, such as $\h\varphi$.

One interesting consequence of this is that, since it can be formulated without any choice of spin structure, the $\Spin(10)$ grand unified theory can conceivably arise as a critical point
in a purely bosonic theory.   Upon symmetry breaking to a subgroup that does not contain the $2\pi$ rotation in $\Spin(10)$ -- such as SU(5), another standard gauge group for
Grand Unification in particle physics -- a dynamical spin structure is generated.  

\subsection{Time-Reversal Symmetry}\label{timereversal}

Another generalization is to incorporate time-reversal symmetry.   The  basic issue here is to ask what time-reversal properties are possessed by a four-dimensional
SU(2) gauge theory with fermions consisting of, for example, a single multiplet with some half-integral isospin $j$.  Let us suppose that we want a time-reversal symmetry 
that commutes with SU(2). To construct a time-reversal invariant theory with a single isospin $j$ multiplet, what we need is then a four-dimensional pseudoreal spinor
representation of either $\Pin^+(1,3)$ or $\Pin^-(1,3)$.    If the four-dimensional spinor representation of one of these groups is pseudoreal, then upon taking the tensor
product with the isospin $j$ representation of SU(2), we will get a real representation of $\Pin^\pm(1,3)\times \SU$ or of
$(\Pin^\pm(1,3) \times\SU)/\Z_2$  (depending on which version of the theory one wants to consider).   This representation will have dimension $4(2j+1)$,  as is appropriate
to describe a single multiplet of fermions of isospin $j$.

To describe $\Pin^+(1,3)$, start with a Clifford algebra of signature $-+++$, 
that is a Clifford algebra with
\be\label{zocc}\{\gamma_\mu,\gamma_\nu\}=2\eta_{\mu\nu},~~~~\eta_{\mu\nu}=\mathrm{diag}(-1,1,1,1).\ee
This Clifford algebra has a representation by $4\times 4$ real matrices.  So the four-dimensional spinor representation of $\Pin^+(1,3)$ is real.
Upon taking the tensor product with the isospin $j$ representation of SU(2), for half-integral $j$, we get a pseudoreal representation of 
$(\Pin^+(1,3) \times\SU)/\Z_2$, so that with $\Pin^+(1,3)\times \SU)/\Z_2$ symmetry, it is not possible to have a single
fermion multiplet of isospin $j$.

On the other hand, the spinor representation of $\Pin^-(1,3)$ is a four-dimensional space on which acts a Clifford algebra of signature $+---$, that is, a Clifford
algebra with
\be\label{zotcc}\{\gamma_\mu,\gamma_\nu\}=-2\eta_{\mu\nu}.\ee
This algebra cannot be represented by $4\times 4$ real matrices, so the four-dimensional spinor representation of $\Pin^-(1,3)$ is
not real.   Rather it is pseudoreal, so $(\Pin^-(1,3)\times \SU)/\Z_2$ has a real representation of the desired type.

Once we generalize the anomalous SU(2) gauge theories studied in this paper to be time-reversal invariant, the anomalies that we have studied will have to generalize.
After all, we are free to forget time-reversal symmetry and just look at these theories on an orientable manifold; the anomalies that we have found cannot disappear because
we have learned that the theory actually has time-reversal symmetry.  

So the discussion in this paper will generalize to $\pin^-$ manifolds with SU(2) gauge fields, and more generally to manifolds with
$(\Pin^-(1,3)\times \SU)/\Z_2$ structure.  The mod 2 index in five dimensions generalizes nicely in this situation.   A few details in the presentation
need some modification.  The obstruction to a $\pin^-$ structure is $w_2(TM)+w_1(TM)^2$, so this will play the role that was played by $w_2(TM)$ in some statements.
In a theory that is required to be time-reversal invariant, the $\theta$ parameter is no longer an arbitrary angle.  Rather, with the normalization that we used in section \ref{Four},
its allowed values are $0,\pi$ (or  $0,\pi/2$ in the $\spinc$ case).

\subsection{Five Dimensions}

Finally, we can consider an $\SU$ gauge theory in five dimensions.   The spinor representation of $\Spin(1,4)$ is pseudoreal, so it is possible in five dimensions
to consider a single multiplet of fermions in a pseudoreal representation, say the isospin $j$ representation of $\SU$ for some half-integer $j$.  (This was discussed in
section \ref{preliminaries}.)  Upon dimensional reduction to four dimensions, 
this theory reduces to the four-dimensional
theory with a single multiplet of fermions with isospin $j$.   This class of four-dimensional theories has anomalies that were studied in section \ref{Two}.   As we will now explain,
 the five-dimensional
version of the theories has very similar anomalies.  

In fact, the examples that we considered in section \ref{Two} to exhibit anomalies in four dimensions can be reinterpreted to exhibit anomalies in five dimensions.
Consider the five-manifold $Y_1=S^4\times S^1$ with an SU(2) instanton bundle of instanton number 1 on the $S^4$ and gauge fields and spin structure pulled back from $S^4$.
The mod 2 index of an isospin 1/2 fermion on this five-manifold is nonzero, as discussed in section \ref{review}.   We interpreted this  nonzero index
in our earlier discussion as a five-dimensional
invariant that measures an anomaly in a four-dimensional theory.  But the same calculation has a direct interpretation in five dimensions:  having an odd number of fermion zero-modes
in this example shows that the operator $(-1)^F$ can be anomalous in a five-dimensional theory with a single fermion multiplet of $j=1/2$.   Thus this theory is anomalous.   (This anomaly
was pointed out in \cite{IMS}.  A certain string theory construction with gauge symmetry Sp(2N) in five dimensions
 always gives an even number of fermion mutliplets in the fundamental representation \cite{Z};
this reflects the fact that the case with an odd number of such multiplets is anomalous.)

The new SU(2) anomaly similarly has an analog in five dimensions.   We simply consider the five-manifold $Y_2=\CP^2\rtimes \S^1$ with the $(\Spin(5)\times \SU)/\Z_2$ structure
that was exploited in section \ref{Two}.  For this $(\Spin(5)\times \SU)/\Z_2$ structure, the mod 2 index of the Dirac operator in the $j=3/2$ representation of SU(2) is nonzero.
In section \ref{fived}, we interpreted this as a five-dimensional invariant that diagnoses a new SU(2) anomaly in four dimensions.   But the same nonzero mod 2 index represents
an anomaly in $(-1)^F$ in a five-dimensional theory with a single fermion of isospin 3/2.  Thus also the new SU(2) anomaly has a five-dimensional analog.  

The six-dimensional cobordism invariant that detects the old and new SU(2) anomalies in five dimensions is simply the mod 2 index of a chiral six-dimensional Dirac operator
acting on fermions in the isospin 1/2 or isospin 3/2 representation of SU(2).  (Here by a chiral Dirac operator in six dimensions, we mean a Dirac operator acting on a spinor
field of one definite chirality.)    Following the logic of section \ref{Two}, as an example of six-manifolds that detect the  old and new SU(2) anomalies, we can just take products
$Y_1\times S^1$ or $Y_2\times S^1$, with metrics, gauge fields, and spin structures all pulled back from $Y_1$ or $Y_2$.

\vskip 1cm \noindent
{\it Acknowledgements}  We thank C. Cordova for helpful remarks.  JW gratefully acknowledges support from  a Corning Glass Works
Foundation Fellowship and  from NSF Grant  PHY-1606531. XGW is partially supported by NSF grants DMR-1506475 and DMS-1664412.  EW is partially supported by NSF Grant PHY-1606531.

\end{document}